\documentclass[12pt,preprint]{aastex}
\usepackage{epsf}

\def\nbody{$n$-body}
\def\deg{\ifmmode {^\circ}\else {$^\circ$}\fi}
\def\degree{\ifmmode {^\circ}\else {$^\circ$}\fi}
\def\mum{\ifmmode {\rm \,\mu {\rm m}}\else $\rm \,\mu {\rm m}$\fi}
\def\arcsec{\ifmmode ^{\prime \prime}\else $^{\prime \prime}$\fi}

\def\inch{\ifmmode ^{\prime \prime}\else $^{\prime \prime}$\fi}
\def\gs{\ifmmode {{\rm g~s^{-1}}}\else ${\rm g~s^{-1}}$\fi}
\def\msunyr{\ifmmode {M_{\odot}~{\rm yr^{-1}}}\else $M_{\odot}~{\rm yr^{-1}}$\fi}
\def\msun{\ifmmode {M_{\odot}}\else $M_{\odot}$\fi}
\def\rsun{\ifmmode {R_{\odot}}\else $R_{\odot}$\fi}
\def\lsun{\ifmmode {L_{\odot}}\else $L_{\odot}$\fi}
\def\mstar{\ifmmode {M_{\star}}\else $M_{\star}$\fi}
\def\rstar{\ifmmode {R_{\star}}\else $R_{\star}$\fi}
\def\tstar{\ifmmode {T_{\star}}\else $T_{\star}$\fi}
\def\lstar{\ifmmode {L_{\star}}\else $L_{\star}$\fi}
\def\mwd{\ifmmode {M_{wd}}\else $M_{wd}$\fi}
\def\rwd{\ifmmode {R_{wd}}\else $R_{wd}$\fi}
\def\twd{\ifmmode {T_{wd}}\else $T_{wd}$\fi}
\def\lwd{\ifmmode {L_{wd}}\else $L_{wd}$\fi}
\def\md{\ifmmode {M_d}\else $M_d$\fi}
\def\ld{\ifmmode {L_d}\else $L_d$\fi}
\def\ad{\ifmmode A_d\else $A_d$\fi}
\def\ldlwd{\ifmmode L_d / L_{wd}\else $L_d / L_{wd}$\fi}
\def\ldlstar{\ifmmode L_d / L_\star\else $L_d / L_{\star}$\fi}
\def\rearth{\ifmmode {\rm R_{\oplus}}\else $\rm R_{\oplus}$\fi}
\def\mearth{\ifmmode {\rm M_{\oplus}}\else $\rm M_{\oplus}$\fi}
\def\qdstar{\ifmmode Q_D^\star\else $Q_D^\star$\fi}
\def\vsqd{\ifmmode v^2 / Q_D^\star\else $v^2 / Q_D^\star$\fi}
\def\kms{\ifmmode {\rm km~s^{-1}}\else $\rm km~s^{-1}$\fi}
\def\ms{\ifmmode {\rm m~s^{-1}}\else $\rm m~s^{-1}$\fi}
\def\vrel{\ifmmode v_{rel}\else $v_{rel}$\fi}
\def\mdot{\ifmmode \dot{M}\else $\dot{M}$\fi}
\def\mdotz{\ifmmode \dot{M}_0\else $\dot{M}_0$\fi}
\def\mesc{\ifmmode m_{esc}\else $m_{esc}$\fi}
\def\rmin{\ifmmode r_{min}\else $r_{min}$\fi}
\def\rmax{\ifmmode r_{max}\else $r_{max}$\fi}
\def\xmax{\ifmmode x_{max}\else $x_{max}$\fi}
\def\mmin{\ifmmode m_{min}\else $m_{min}$\fi}
\def\mmax{\ifmmode m_{max}\else $m_{max}$\fi}
\def\rmind{\ifmmode r_{min,d}\else $r_{min,d}$\fi}
\def\rmaxd{\ifmmode r_{max,d}\else $r_{max,d}$\fi}
\def\mmaxd{\ifmmode m_{max,d}\else $m_{max,d}$\fi}
\def\vrad{\ifmmode v_{rad}\else $v_{rad}$\fi}
\def\qz{\ifmmode q_{0}\else $q_{0}$\fi}
\def\qi{\ifmmode q_{i}\else $q_{i}$\fi}
\def\ql{\ifmmode q_{l}\else $q_{l}$\fi}
\def\qs{\ifmmode q_{s}\else $q_{s}$\fi}
\def\vhill{\ifmmode v_H\else $r_H$\fi}
\def\rhill{\ifmmode r_H\else $r_H$\fi}
\def\Rhill{\ifmmode R_H\else $R_H$\fi}
\def\rbrk{\ifmmode r_{brk}\else $r_{brk}$\fi}
\def\rdamp{\ifmmode r_{damp}\else $r_{damp}$\fi}
\def\rin{\ifmmode r_{in}\else $r_{in}$\fi}
\def\rout{\ifmmode r_{out}\else $r_{out}$\fi}
\def\tin{\ifmmode t_{in}\else $t_{in}$\fi}
\def\tout{\ifmmode t_{out}\else $t_{out}$\fi}
\def\ain{\ifmmode a_{in}\else $a_{in}$\fi}
\def\aout{\ifmmode a_{out}\else $a_{out}$\fi}
\def\r0{\ifmmode r_{0}\else $r_{0}$\fi}
\def\R0{\ifmmode R_{0}\else $R_{0}$\fi}
\def\m0{\ifmmode m_{0}\else $m_{0}$\fi}
\def\M0{\ifmmode M_{0}\else $M_{0}$\fi}
\def\xm{\ifmmode x_{m}\else $x_{m}$\fi}
\def\sigz{\ifmmode \Sigma_0\else $\Sigma_0$\fi}
\def\ergg{\ifmmode {\rm erg~g^{-1}}\else ${\rm erg~g^{-1}}$\fi}
\def\gyr{\ifmmode {\rm g~yr^{-1}}\else ${\rm g~yr^{-1}}$\fi}
\def\cms{\ifmmode {\rm cm~s^{-1}}\else ${\rm cm~s^{-1}}$\fi}
\def\gcms{\ifmmode {\rm g~cm^{-2}}\else $\rm g~cm^{-2}$\fi}
\def\gcmc{\ifmmode {\rm g~cm^{-3}}\else $\rm g~cm^{-3}$\fi}
\def\atil{\ifmmode {\tilde{a}}\else $\tilde{a}$\fi}
\def\ttil{\ifmmode {\tilde{t}}\else $\tilde{t}$\fi}
\def\sqrttt{\ifmmode {\tilde{t}^{1/2}}\else $\tilde{t}^{1/2}$\fi}

\def\orch{{\it Orchestra}}
\def\voy{{\it Voyager}}
\def\cas{{\it Cassini}}
\def\nh{{\it New Horizons}}
\def\pc{Pluto--Charon}
\def\mp{\ifmmode M_P\else $M_P$\fi}
\def\mc{\ifmmode M_C\else $M_C$\fi}
\def\mh{\ifmmode M_H\else $M_H$\fi}
\def\mk{\ifmmode M_K\else $M_K$\fi}
\def\ms{\ifmmode M_S\else $M_S$\fi}
\def\mn{\ifmmode M_N\else $M_N$\fi}
\def\rp{\ifmmode R_P\else $R_P$\fi}
\def\rc{\ifmmode R_C\else $R_C$\fi}
\def\apc{\ifmmode a_{PC}\else $a_{PC}$\fi}

\begin{document}

\title{A Pluto--Charon Sonata: Dynamical Limits on the Masses of the Small Satellites}

\author{Scott J. Kenyon}
\affil{Smithsonian Astrophysical Observatory,
60 Garden Street, Cambridge, MA 02138}
\email{e-mail: skenyon@cfa.harvard.edu}

\author{Benjamin C. Bromley}
\affil{Department of Physics \& Astronomy, University of Utah,
201 JFB, Salt Lake City, UT 84112}
\email{e-mail: bromley@physics.utah.edu}

\begin{abstract}

During 2005--2012, images from Hubble Space Telescope (HST) revealed four moons
orbiting Pluto-Charon \citep{weaver2006,showalter2011,showalter2012}.  Although
their orbits and geometric shapes are well-known, the 2$\sigma$ uncertainties in
the masses of the two largest satellites -- Nix and Hydra -- are comparable to
their HST masses \citep{brozovic2015,showalter2015,weaver2016}. Remarkably,
gravitational $n$-body computer calculations of the long-term system stability
on 0.1--1~Gyr time scales place much tighter constraints on the masses of Nix and
Hydra, with upper limits $\sim$ 10\% larger than the HST mass.  Constraints on the
mass density using size measurements from \nh\ suggest Nix and Hydra formed in
icier material than Pluto and Charon.

\end{abstract}


\section{INTRODUCTION}
\label{sec: intro}

Throughout the solar system, the dynamical architecture of the systems of planets,
moons, and smaller objects provides clues to our origin. The internal structure of
the asteroid belt and the small mass of Mars constrain the formation of Jupiter
and the disappearance of the protosolar nebula \citep[e.g.,][]{walsh2011,
izidoro2014,brasser2016,bk2017,clement2019}. Different classes of Kuiper belt 
objects just beyond the orbit of Neptune hold traces of the early orbital 
evolution of the gas giant planets \citep[r.g.,][]{malhotra1993,ida2000b,
levison2003b, gomes2004a,tsig2005,dawson2012,holman2018}.
Sedna and other transneptunian objects may point to a ninth planet orbiting the
Sun at distances more than ten times beyond the orbit of Neptune \citep{trujillo2014,
batygin2016,brown2016,sheppard2016b,becker2018,sheppard2018,brown2019}.

The binary dwarf planet Pluto-Charon illustrates the chaotic early history of
the solar system \citep{mckinnon1989,canup2005,canup2011,kb2014b,walsh2015,
quill2017,mckinnon2017,stern2018,woo2018}.  
In current theory, Pluto and Charon form in separate locations outside the
orbits of the gas giants.  Charon then suffers a glancing collision with Pluto
and remains bound. Icy debris from the collision coalesces into the small satellites.
Because the debris is mostly ice, the density of the satellites should be smaller
than the density of either Pluto or Charon. Current observational limits on the
satellite masses and mass densities are poor \citep{weaver2016}.

Here, we place better limits on the satellite masses with $n$-body calculations.
Together with measured sizes from the \nh\ mission, we then derive the mass
density and test the formation theories.  Along with the high measured albedos
from \nh, the new estimates of the mass density support theoretical models where
the satellites form in icy material left over from the collision of Charon with
Pluto. Other options -- such as capture from the Kuiper belt -- are inconsistent
with the low density and high albedo \citep{lith2008b,kb2014b,walsh2015,weaver2016}.

This analysis demonstrates the power of dynamical $n$-body calculations combined
with accurate observations of orbits, shapes, and sizes. Aside from improving our
understanding of the environment in which the satellites formed, the mass estimates
from the $n$-body calculations provide new tests of theories for circumbinary dynamics
and establish clear targets for the next generation of formation calculations.

We begin with an observational (\S\ref{sec: obs}) and a theoretical
(\S\ref{sec: theory}) summary of the \pc\ system. We then describe the
\nbody\ code (\S\ref{sec: code}), starting conditions for each calculation
(\S\ref{sec: start}), several numerical tests (\S\ref{sec: tests}), our
formalism for measuring the stability of the satellite system
(\S\ref{sec: stable}), and the results of the calculations (\ref{sec: results}).
New limits on the masses of Nix and Hydra allow robust estimates of the
mass density for each satellite (\S\ref{sec: density}). After discussing
the significance of our analysis (\S\ref{sec: disc}), we conclude with a
brief summary (\S\ref{sec: summary}).

\vskip 6ex
\section{OBSERVATIONAL BACKGROUND}
\label{sec: obs}

Charon and Pluto orbit the system barycenter every 6.387 days. Tidal forces maintain
rotational periods equal to the orbital period \citep{christy1978,buie1997,buie2006,
buie2012}.
For an adopted gravitational constant $G = 6.67408 \times 10^{-8}$, Pluto has a
mass $m_P = 1.303 \times 10^{25}$~g, radius $R_P$ = 1188.3~km, mean density
$\rho_P$ = 1.854~g~cm$^{-3}$, and oblateness $f_P \le$ 0.006.  Charon is somewhat
smaller, with mass $m_C = 1.586 \times 10^{24}$~g, radius $R_C$ = 606~km,
mean density $\rho_C$ = 1.702~g~cm$^{-3}$, and oblateness $f_C \le$ 0.005.
With a center-of-mass outside Pluto's surface, the pair is a true binary planet
\citep{stern2015,nimmo2017}.

The four small circumbinary satellites lie on nearly circular orbits (orbital
eccentricity $e \le$ 0.006) inclined at less than 1 degree from the Pluto-Charon
orbital plane (Table~\ref{tab: sats}). The orbital periods are 3.16 (Styx),
3.89 (Nix), 5.04 (Kerberos), and 5.98 (Hydra) times the orbital period of
Pluto-Charon.  Three moons -- Styx, Nix, and Hydra -- may lie in a three body
resonance where the ratio of synodic periods is $3 S_{NH}$ = $2 S_{SN}$
\citep{showalter2015}.  HST and \nh\ data also demonstrate that the satellites
rotate chaotically, with periods much shorter than their orbital periods
\citep{showalter2015,weaver2016}.

Despite sensitive imaging observations, \nh\ did not detect any additional
small satellites \citep{weaver2016}. The upper limit on the radius for other
moons with orbital semimajor axis $a \le$ 80,000~km from the system barycenter
(roughly 1.25 times the orbital distance of Hydra) is $R \le$ 1.7~km, nearly
3 times smaller than the spherical radius of Styx. Upper limits on the existence
of rings or other debris in the system is also severe: an unseen ring or cloud
must have an optical depth smaller than $10^{-8}$ \citep{lauer2018}.

The \nh\ images confirm irregular, oblong shapes for each satellite
\citep{weaver2016}, with approximate dimensions (in km) of
$16 \times 9 \times 8$ (Styx), $50 \times 35 \times 33$ (Nix),
$19 \times 10 \times 9$ (Kerberos), and $65 \times 45 \times 25$ (Hydra).
Assuming their shapes are triaxial ellipsoids, the equivalent spherical radii
are 5.2 km (Styx), 19.3~km (Nix), 6.0~km (Kerberos), and 20.9~km (Hydra).
For the HST masses, mean densities are $\rho \approx $  6.5~g~cm$^{-3}$ (Styx),
1.49~g~cm$^{-3}$ (Nix), 18.2~g~cm$^{-3}$ (Kerberos), and 1.26~g~cm$^{-3}$ (Hydra).
The densities for Nix and Hydra are somewhat smaller than the mass density
of either Pluto or Charon, consistent with an icier composition.  The
3$\sigma$ upper limits on the sizes of Styx and Kerberos, $R_S \le$ 10~km
and $R_K \le$ 12~km, allow mass densities more similar to Pluto/Charon and
Nix/Hydra: $\rho_S \ge$ 0.95~g~cm$^{-3}$ and $\rho_K \ge$ 2.1~g~cm$^{-3}$.

To provide an alternative to the `heavy' satellite system with the nominal HST
masses, we consider a `light' system with more physically plausible densities
for Styx and Kerberos.  Setting $\rho \approx$ 1~g~cm$^{-3}$ yields
$M_S \approx 6 \times 10^{17}$~g (Styx) and $M_K \approx 10^{18}$~g (Kerberos).
These masses are consistent with HST masses at the 1$\sigma$ level for Styx and
at the 2$\sigma$ level for Kerberos.

Before the discovery of Styx, $N$-body simulations of the orbital stability of a
massless Kerberos implied relatively low masses for Nix and Hydra, no more than
10\% (Nix) or 90\% (Hydra) larger than the HST masses in Table~\ref{tab: sats}
\citep{youdin2012}. For mass density $\rho$ = 1~g~cm$^{-3}$, the predicted
reflectivity (albedo) then exceeds $A$ = 0.3.  The observed albedos confirm
this prediction \citep{weaver2016}. Hydra ($A$ = 0.83) is an almost perfect
reflector; Styx ($A$ = 0.65), Nix ($A$ = 0.56) and Kerberos ($A$ = 0.56)
are also highly reflective.

Activity of the three body resonance among Styx, Nix, and Hydra also constrains
the masses of Nix and Hydra \citep{showalter2015}. Adopting the HST mass
for Kerberos in Table~\ref{tab: sats}, the resonance is inactive if the masses
of Nix and Hydra are less than roughly 1.5 times their HST masses.  For
somewhat larger masses, the resonance is active.  Thus, there is some tension
between the masses needed for an active resonance and those consistent with
previous $N$-body simulations or implied by measured physical sizes.  Our
$N$-body calculations rule out the large masses for Nix and Hydra required
for an active resonance in a heavy satellite system.

\vskip 6ex
\section{THEORETICAL BACKGROUND}
\label{sec: theory}

In any system of satellites orbiting a central binary system, the Hill radius
$R_H$ establishes a boundary where the gravity of the satellite dominates the
gravity of the central binary.  The radius of the Hill sphere for a single
satellite orbiting Pluto-Charon is
\begin{equation}
R_H = (m_i / 3 (m_P + m_C))^{1/3} ~ a_i ~ ,
\label{eq: rhill1}
\end{equation}
where $a_i$ is the orbital distance of a satellite of mass $m_i$. Within a
roughly spherical volume of radius $r = R_H$ centered on the satellite, material
comoving with the satellite is bound to the satellite. Every satellite orbiting
Pluto-Charon has its own `Hill sphere'. Material orbiting Pluto-Charon outside
these Hill spheres is bound to Pluto-Charon \citep{nagy2006,suli2009,pires2011,
smullen2017}. Table~\ref{tab: sats} lists Hill radii for each satellite.

For the four small satellites of Pluto-Charon, the relative sizes of the
mutual Hill radii set limits on the orbital stability of the system.  Taking
adjacent satellites in pairs, the mutual Hill radius is
\begin{equation}
R_{H,ij} = \bar{a} ( (m_i + m_j) / 3 (m_P + m_C) )^{1/3} ~ ,
\label{eq: rhill2}
\end{equation}
where $\bar{a}$ = $(a_i + a_j) / 2$ is the average of the two semimajor axes and
$m_i + m_j$ is the sum of the masses. Defining a normalized orbital separation
$K_{ij} = |a_i - a_j| / R_{H,ij} $, the physical parameters for the heavy
satellite system in Table~\ref{tab: sats} imply $K_{SN}$ = 12 for Styx--Nix,
$K_{NK}$ = 16 for Nix--Kerberos, and $K_{KH}$ = 10 for Kerberos--Hydra. The
masses for the light satellite system yield similar values:
$K_{SN}$ = 13 for Styx--Nix, $K_{NK}$ = 17 for Nix--Kerberos, and
$K_{KH}$ = 11 for Kerberos--Hydra.

To put the orbital periods and mutual Hill separations in context, we compare
with the giant planets in the solar system and the Galilean moons of Jupiter.
Relative to the orbital period of Jupiter $P_J$, the four gas giants have
orbital periods $P_S \approx$ 2.5 $P_J$, $P_U \approx$ 7 $P_J$, and
$P_N \approx$ 14; the mutual Hill separations are $K_{JS}$ = 5.3
(Jupiter--Saturn), $K_{SU}$ = 9.7 (Saturn--Uranus), and $K_{UN}$ = 9.6
(Uranus--Neptune). The inner three Galilean satellites (Io, Europa, and Ganymede)
are in orbital resonance where the orbital period of Europa (Ganymede) is twice
(four times) the orbital period of Io around Jupiter. The period of Callisto is
9.4 times Europa's period. These moons have nearly identical mutual Hill separations,
$K_{IE} \approx$ 11 (Io--Europa), $K_{EG} \approx$ 10 (Europa--Ganymede), and
$K_{GC} \approx$ 11 (Ganymede--Callisto). The \pc\ satellites are not quite as
closely packed as the giant planets, but have similar mutual Hill separations
as the Galilean moons.

Dynamical theory places strong constraints on the mutual Hill separations
for systems of planets or satellites orbiting a single central object
\citep{wisdom1980,petit1986,gladman1993,deck2013}. For pairs of equal mass
satellites, $K \ge$ 3.5 ensures stability. The Jupiter--Saturn system
easily satisfies this constraint.  For 3 or more satellites, the minimum $K$
is sensitive to the masses of the satellites relative to each other and to
the mass of the central object \citep{chambers1996,smith2009,fang2013,
kratter2014,fabrycky2014,mahajan2014,pu2015,morrison2016,obertas2017}. 
$K \approx$ 10--12 is typical. The outer three gas giants and the Galilean 
satellites satisfy this constraint.

When the central object is a binary, the set of stable orbits is much smaller.
Coplanar prograde, circumbinary orbits must lie outside a critical semimajor
axis, $a_c \approx 1.7 - 2 ~ a_{pc}$, where $a_{PC}$ is the semimajor axis of
the Pluto-Charon binary \citep{holman1999,doolin2011,chavez2015,li2016,
quarles2018b,lam2018,kb2019}. The orbital semimajor axis of Styx, 
$a_S \approx 2.4 ~ a_{PC}$, meets this criterion.  Although there are 
limited analyses of the stability of multi-planet systems in binary 
systems \citep{kratter2014,marzari2016}, stability with 
$K \le$ 10--12 seems unlikely.

Dynamical theory suggests that the four satellites are probably unstable if
all of the HST masses are correct. With $K_{SN}$ = 12 for the Styx--Nix pair
and $K_{NK}$ = 16 for Nix--Kerberos, the Styx--Nix and Nix--Kerberos pairs
are safely stable on their own.  With $K_{KH}$ = 10, however, the
Kerberos--Hydra pair is at the nominal stability limit for circumbinary
satellites \citep{kratter2014,marzari2016}. With three pairs of 
satellites at or close to the nominal stability limit, a stable system 
probably requires smaller satellite masses.

Dynamical theory is more ambiguous for the light system. As in the heavy
system, the Styx--Nix and Nix--Kerberos pairs are well within the stability
limits.  With $K_{KH}$ = 11, the Kerberos--Hydra pair is probably also
stable on its own. However, the extra gravity of Nix might be sufficient
to push the system beyond the stability limit.  Our goal is to test these
predictions of dynamical theory for light and heavy satellite systems.

\vskip 6ex
\section{CALCULATIONS}

\vskip 2ex
\subsection{$N$-body Code}
\label{sec: code}

To explore the long-term stability of the Pluto-Charon satellite system,
we perform numerical calculations with a gravitational $n$-body code which
integrates the orbits of Pluto, Charon, and the four smaller satellites in
response to their mutual gravitational interactions.  Our $N$-body code,
\orch, employs an adaptive sixth-order accurate algorithm based on either
Richardson extrapolation \citep{bk2006} or a symplectic method
\citep{yoshida1990,wisdom1991,saha1992}.  The code calculates gravitational
forces by direct summation and evolves particles accordingly in the
center-of-mass frame.  Aside from passing a stringent set of dynamical tests
and benchmarks \citep{dunc1998,bk2006}, we have used the code to simulate
scattering of super-Earths by growing gas giants \citep{bk2011a}, migration
through planetesimal disks \citep{bk2011b} and Saturn's rings \citep{bk2013},
the formation of Pluto's small satellites \citep{kb2014b}, the circularization
of the orbits of planet scattered into the outer solar system \citep{bk2014,
bk2016}, and the potential for discovering other satellites in the Pluto-Charon
system \citep{kb2019}. In several of these studies, we describe additional
tests of the algorithm.

In these calculations, we do not consider tidal or radiation pressure forces
on the satellites \citep[e.g.,][]{burns1979,hamilton1992,poppe2011,pires2013,
quill2017}. Although radiation pressure forces are significant on dust grains,
satellites with sizes similar to Styx and Kerberos are unaffected. As long as
the orbit of the central binary remains fixed, tidal forces should have little
impact on the orbits of the small satellites.

During the symplectic integrations, there is no attempt to resolve collisions
between the small satellites or between an ejected satellite and Pluto or
Charon. Satellites passing too close to another massive object in the system
are simply ejected. In the adaptive integrator, the code changes the length
of timesteps to resolve collisions.  In agreement with previous results
\citep{sutherland2016,smullen2016,smullen2017}, small satellites are always
ejected from the system and never collide with other small satellites,
Charon, or Pluto.

Previous \nbody\ calculations suggest that the orbits of the small satellites
are too far inside the Hill sphere of Pluto to require including the gravity
of the Sun or major planets in the integrations \citep{michaely2017}. For
reference, the radius of the Pluto-Charon Hill sphere is
$R_{H,PC} \approx 8 \times 10^6$~km. In Hill units, the radius of Hydra's orbit,
$a_H / R_{H,PC} \approx$ 0.008, is well inside the Hill sphere and fairly immune
from the gravity of the Sun.  For most of the calculations described in this
paper, the \nbody\ code follows the orbits of \pc\ and the four small satellites
without any contribution from the gravity of the Sun or major planets. As a test,
some simulations include the Sun and major planets; however, these calculations
yield results identical to those without extra sources of gravity.

Throughout the \nbody\ calculations, we record the 6D cartesian phase space
variables, the orbital eccentricity $e$, and the orbital inclination $\imath$
at the end of selected time steps.  Over total integration times as long as
0.1--1~Gyr, a typical calculation has 30,000 to more than 100,000 of these
`snapshots' of the satellite positions, velocities, and orbital parameters
at machine precision.  To avoid unwieldy data sets, we make no attempt to 
record satellite positions during each orbit. Within the circumbinary 
environment of \pc, satellite orbits precess on time scales ranging from 
1.2~yr for Styx to 2.8~yr for Hydra \citep[e.g.,][]{lee2006,leung2013,bk2015a}.  
For any calculation, the ensemble of snapshots is insufficient to track the 
precession of the small satellites.

On the NASA `discover' cluster, 24~hr integrations on a single processor advance
the satellite system $\sim$ 4.3~Myr.  We perform 28 calculations per node, with 
each satellite system evolving on one of the 28 cores per node.  To derive results 
for as many sets of initial conditions as possible, the suite of simulations uses 
6--10 nodes each day. In this way, each system advances $\sim$ 125~Myr per month.

\vskip 4ex
\subsection{Starting Conditions}
\label{sec: start}

All calculations begin with the same {\it measured} initial state vector
\citep{brozovic2015} for the 3D cartesian position -- $\vec{r} = (x, y, z)$ --
and velocity -- $ \vec{v} = (v_x, v_y, v_z)$ -- of each component.
Tests with a state vector downloaded from the JPL Horizons
website\footnote{https://ssd.jpl.nasa.gov/horizons.cgi} yield indistinguishable
results.  With no state vector for Pluto available from the HST data
\citep{brozovic2015}, we derive results for two options: setting position
and velocity vectors for Pluto assuming the four small satellites
(i) are massless or (ii) have the nominal masses for the light and
heavy systems in Table~\ref{tab: sats}.  As listed in Table~\ref{tab: start-pos},
the resulting differences in $(x, y, z)$ and $(v_x, v_y, v_z)$ velocity for
Pluto are less than 0.2~km in position and 0.5~cm~s$^{-1}$ in velocity. At
a typical distance of roughly 40~AU, a distance of 1~km subtends an angle
of less than 0.0001 arcsec, which is much smaller than the residuals in the
orbits of any of the satellites.  Several tests demonstrate negligible
differences in outcomes for otherwise identical calculations starting from
the three different starting points in Table~\ref{tab: start-pos}. For
simplicity, we rotate the coordinate system to place the \pc\ orbit in the
$x-y$ plane.

Although all calculations begin with the same initial state vector for Pluto,
Charon, and, the four small satellites, we perform each simulation with slightly
different satellite masses. In some sets of calculations, we multiply the nominal
masses for each circumbinary satellite in Table~\ref{tab: sats} by a factor
$f = n (1 + \delta)$, where $n$ is an integer or simple fraction (e.g., 1.25 or 1.5)
and $\delta$ is a small real number in the range $-$0.01 to 0.01; for a suite of 
calculations with similar $f$, $n$ and $\delta$ are {\it the same for all satellites}. 
Varying $n$ and $\delta$ among the full ensemble of calculations provides a way to 
measure the lifetime of nearly identical configurations with the same starting 
positions, orbital velocities, and $n$.  In other simulations, we multiply the mass 
of a single satellite by a factor $f_i$ and set the masses of the remaining satellites 
at their nominal masses. Instead of conducting many simulations with nearly identical 
$f$, we densely sample a set of real $f_i \le $ 5.

To avoid confusion, we use $f$ as a marker for calculations where we multiply the 
masses of all satellites by a common factor and $f_H$ (Hydra), $f_N$ (Nix), or
$f_K$ (Kerberos) as markers when only one satellite has a mass that differs from 
the nominal masses for a light or a heavy system.

Once a calculation begins, all of the evolutionary sequences follow the same
trend. After a period of relative stability where the orbital parameters of
the system are roughly constant in time, the motion of at least one satellite
begins to deviate from its original orbit.  These deviations grow larger and
larger until the orbits of at least two satellites cross.  Instead of colliding,
at least one satellite -- usually Styx or Kerberos -- is ejected from the system
shortly after the first orbit crossing.  We define the lifetime of the system
$\tau_i$ as the evolution time between the start of a calculation and the moment
when one of the satellites is ejected beyond the \pc\ Hillsphere.  Lifetimes
range from 1--10~yr for very massive (and unlikely) satellite systems to more
than 1~Gyr for heavy or light systems with the nominal masses.  The uncertainty
of the ejection time is less than 0.1\% of $\tau_i$.  When we perform $M$
calculations with nearly identical starting conditions, we adopt $\tau_m$ -- the
median of $M$ different $\tau_i$ -- as the lifetime of the system. For fixed $f$,
the range in $\tau_i$ is $\sim$ a factor of 10.  Within the set of calculations
where we change the mass of only one satellite, we look for trends in $\tau_i$
with $f$.

\vskip 4ex
\subsection{Examples}
\label{sec: tests}

To evolve the orbit of the Pluto-Charon satellites in time, the time step in our
symplectic algorithm is $\Delta t = T_{PC} / N$ where $T_{PC}$ is the orbital period
of the central binary and $N$ is an integer. For any simulation, the total processor 
time is proportional to $N$. To select a value for $N$ which maintains the integrity 
of the solution in a reasonable amount of processor time, \citet{kb2019} consider the 
orbit of an idealized Pluto-Charon binary with the measured masses and orbital 
semimajor axes and orbital eccentricity $10^{-4}$, $10^{-5}$, $10^{-6}$, and $10^{-7}$.
In these tests, there is no indication that the average/median $a$ (and its standard 
deviation or inter-quartile range) or any trend in $a$ or $e$ with time depend on the 
number of steps per binary orbit. However, the ability to maintain the input $e$ 
depends on $N$: large $N$ maintains the lowest $e$ orbits better than smaller $N$ 
\citep{kb2019}.  Faithfully tracking the Pluto-Charon binary requires $N \ge$ 40.

Long ($\ge$ 100~Myr) simulations of Pluto-Charon and the four satellites with
$N$ = 40 yield similar results for trends in $a(t)$ and $e(t)$ of the central
binary. As long as the satellite system remains stable, there is no trend in
$a$ or $e$ of the Pluto-Charon binary.  Over the complete simulation, the
dispersion in $a$ and $e$ for Pluto-Charon is larger when the satellite system
is unstable than in a system with no circumbinary satellites.  Once the orbits
of the small satellites begin to cross, the orbit of Pluto--Charon changes in
proportion to the masses of the satellites: disruption of more massive satellite
systems generates larger changes in $a$ and $e$ for Pluto-Charon.

\begin{figure}
\begin{center}
\includegraphics[width=5.5in]{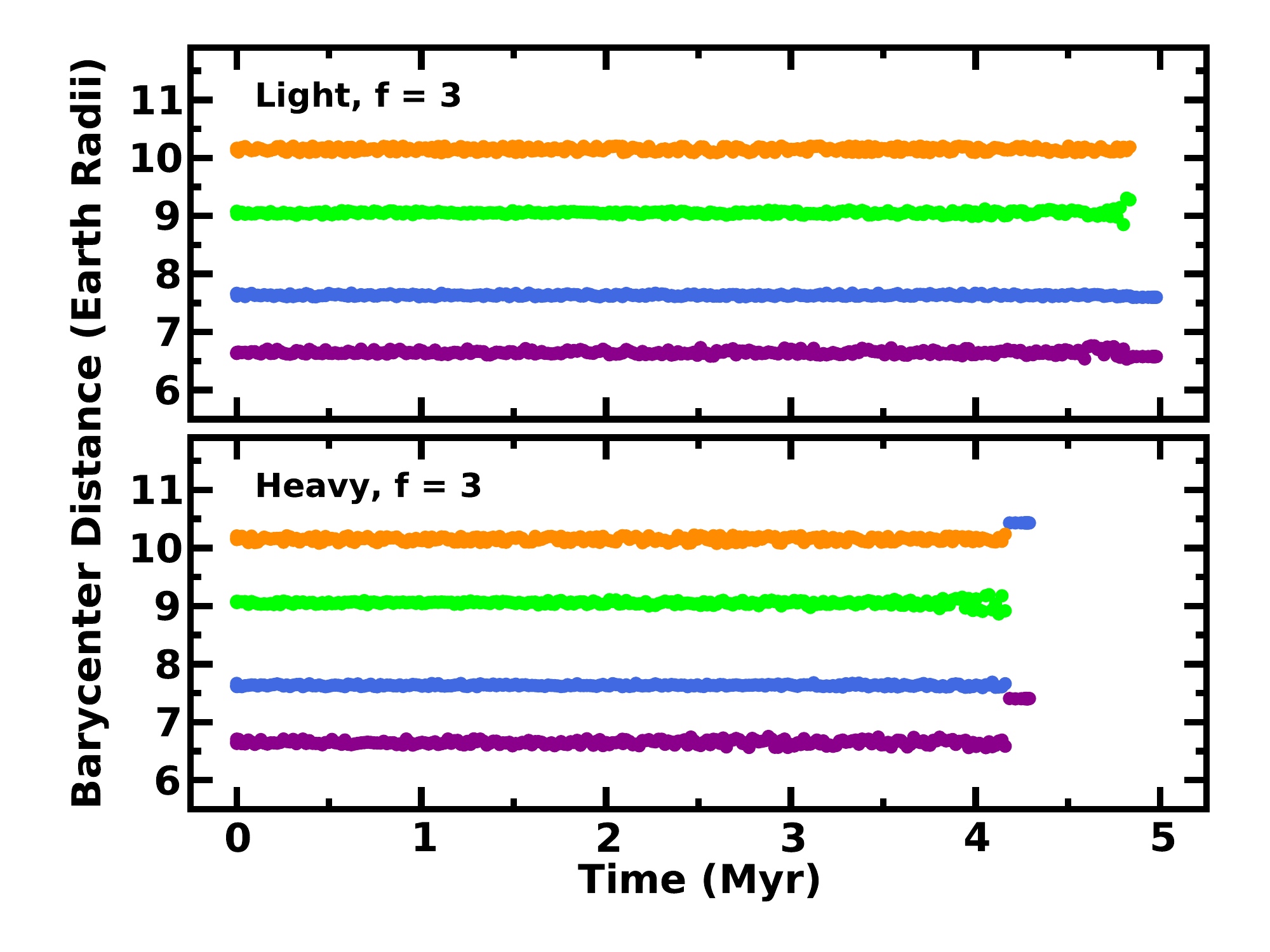}
\end{center}
\vskip -5ex
\caption{
\label{fig: test1}
Satellite ejection in light (upper panel) and heavy (lower panel) systems with
$N$ = 40 and mass ratio $f$ = 3 for all satellites. Symbols plot $r$ the
distance from the barycenter for Styx (purple), Nix (blue), Kerberos (green),
and Hydra (orange) as a function of time.
}
\end{figure}

Fig.~\ref{fig: test1} illustrates the long-term evolution of systems where
$f$ = 3 for all satellites.  In the heavy system (lower panel), the satellites
maintain a constant distance from the system barycenter for roughly 4~Myr.
All of the satellites then develop eccentric orbits ($e$ = 0.06--0.2). Once
Kerberos crosses the orbit of Hydra, both are lost. The remaining satellites
lie on eccentric orbits at larger distances from the system barycenter.

In the light system, Nix and Hydra gradually excite the orbit of Kerberos. However,
Kerberos is not massive enough to modify the orbits of Nix or Hydra. After nearly 5~Myr,
Kerberos crosses the orbit of Hydra, destabilizing the orbits of both satellites. Both
are ejected from the system. Although ejection has limited impact on $e$ for Styx and
Nix, both orbit at somewhat smaller distances from the system barycenter.

Throughout the evolution, the oscillations in the orbital distances and eccentricities for
satellites in the heavy system are larger than those in the light system. With a more
massive Styx and Kerberos, the mutual gravitational interactions are larger in the heavy
system, resulting in larger perturbations of the orbits. Despite this difference,
lifetimes are not significantly different, 4.17~Myr for the heavy system and 4.85~Myr for
the light system. Among all of the calculations for $f$ = 3, the heavy system has a
marginally smaller median lifetime (2.9~Myr) than the light system (3~Myr). These lifetimes
are much shorter than the 4.5~Gyr age of the solar system; thus, these systems are unlikely
proxies for the Pluto-Charon satellite system.

\begin{figure}
\begin{center}
\includegraphics[width=5.5in]{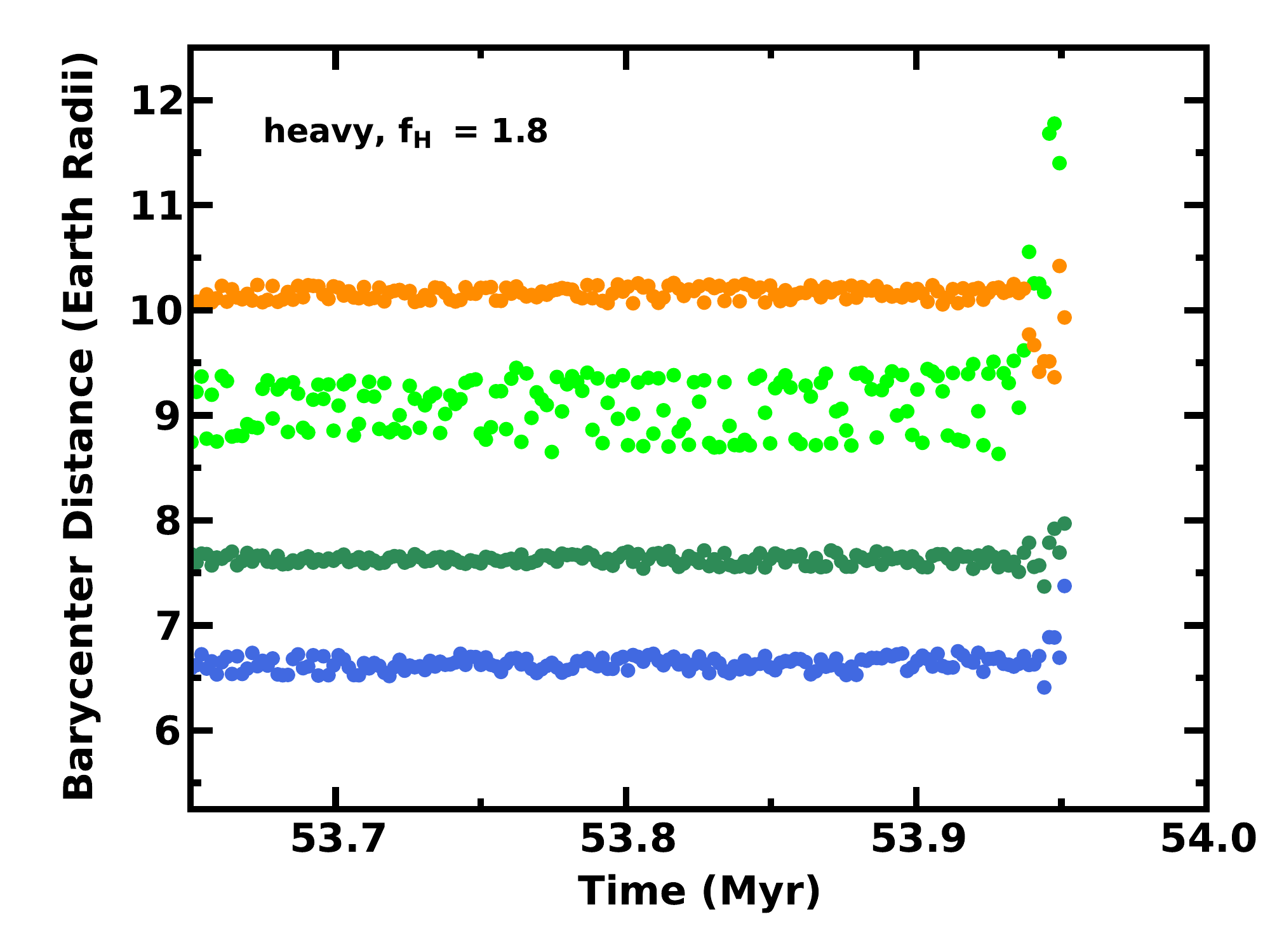}
\end{center}
\vskip -5ex
\caption{
\label{fig: test2}
As in Fig.~\ref{fig: test1} for a system with $N$ = 40 and a massive
Hydra ($f_H$ = 1.8) in a heavy system with nominal masses for the other
satellites.
}
\end{figure}

Fig.~\ref{fig: test2} focuses on the last stages of a calculation with heavy
satellites and $f_H$ = 1.8.  Near the end of the evolution, the orbital distance
of Kerberos (light green points) varies chaotically from roughly 8.5~$R_\oplus$
to nearly 9.5~$R_\oplus$ from the system barycenter.  Because Styx (blue points)
and Nix (dark green points) are much closer to Pluto-Charon than Kerberos, they
are less affected by the large mass of Hydra (orange points). Near the end of
the tracks, Kerberos crosses the path of Hydra, passing outside of Hydra's orbit
and pulling Hydra closer to Pluto-Charon. After reaching an orbital distance of
12~$R_\oplus$ from the barycenter, Kerberos returns to approach Hydra's orbit and
is then ejected.  Hydra returns close to its original orbital distance, but on
an eccentric orbit ($e$ = 0.05). Nix and Styx then lie on slightly wider, more
eccentric orbits. The high eccentricity of Styx ($e$ = 0.14) guarantees that it
will eventually cross the orbit of Nix and be ejected.

\begin{figure}
\begin{center}
\includegraphics[width=5.5in]{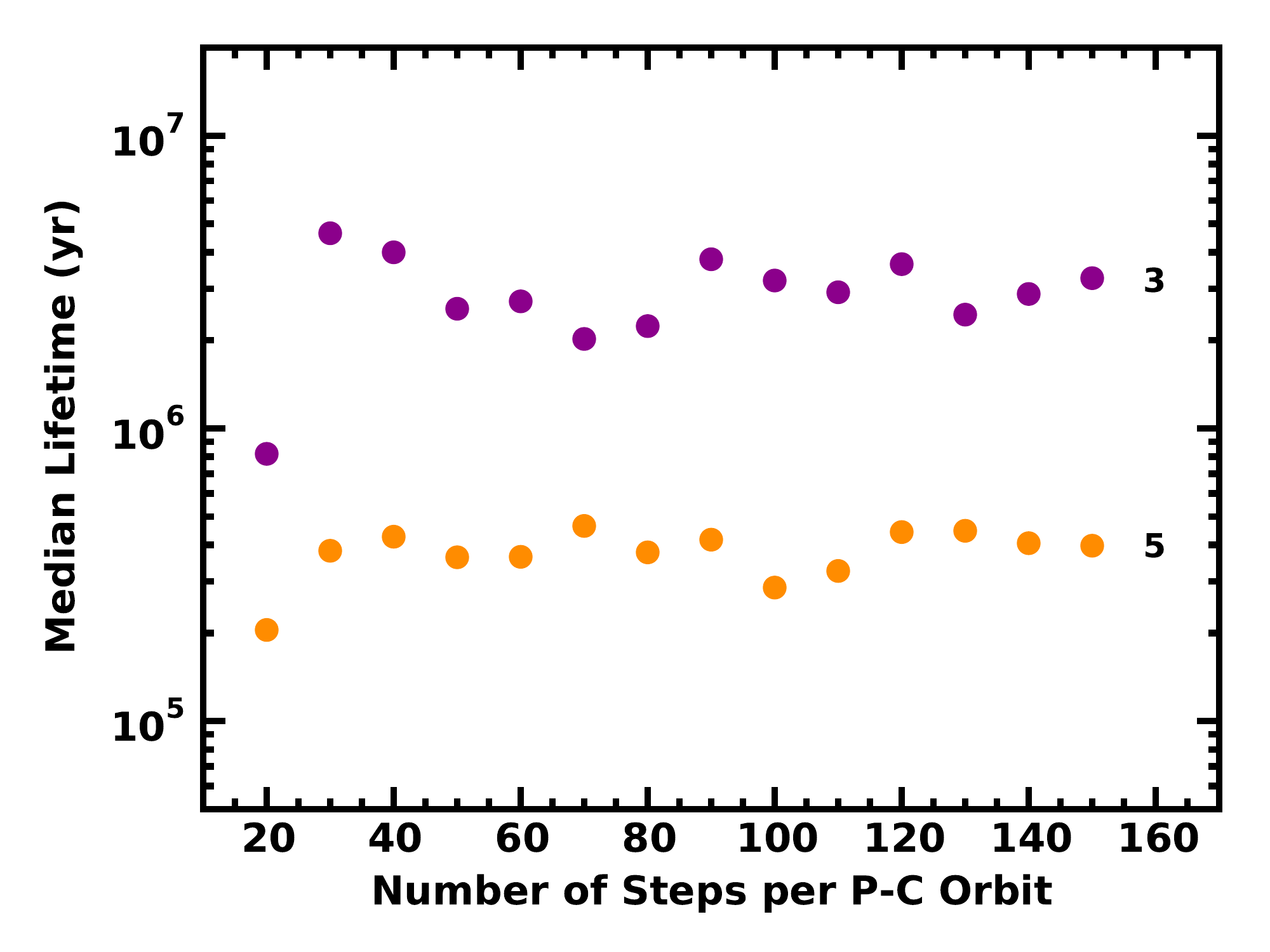}
\end{center}
\vskip -5ex
\caption{
\label{fig: test3}
Median lifetime $\tau_m$ as a function of $N$ from 14--20 calculations of
satellite systems with $f =$ 3 (purple symbols) and $f =$ 5 (orange symbols).  
Calculations with 30 or more steps per orbit yield similar values for $\tau_m$; 
simulations with $N$ = 20 result in much smaller lifetimes.
}
\end{figure}

Fig.~\ref{fig: test3} shows results for the median lifetime of the satellite system
for $N$ = 20--150, $f$ = 3, and $f$ = 5. For each combination of $N$ and $f$,
11--15 calculations yields a robust median; within the 1$\sigma$ dispersion of the
lifetimes, the median and average lifetimes are identical.  For $N$ = 30--150 and
all $f$, the median lifetime is independent of $N$.  When $N$ = 20, the average and
median lifetimes are systematically smaller than calculations with larger $N$.
This pattern persists for larger $f$, albeit with somewhat larger scatter.

\vskip 4ex
\subsection{Stability Considerations}
\label{sec: stable}

In some calculations, none of the four small satellites are ejected after
900--1100~Myr of orbital integration. Completing 4.5~Gyr of integration is
computationally intensive and beyond the scope of our effort.  To judge 
whether these configurations are stable on 4.5~Gyr time scales, we perform 
statistical analyses of satellite orbits in each calculation.

To analyze the 20,000--100,000 snapshots for a simulation with no ejection,
we compute the average distance $r_{avg, i}$ of each satellite from the
barycenter and the average height $z_{avg,i}$ of each satellite from the
plane of the Pluto-Charon binary for all snapshots; we then calculate
$\delta r_i$ = $|r_i - r_{avg_i}|$ and $\delta z_i$ = $|z_i - z_{avg_i}|$.
From estimates of the linear correlation coefficient (Pearson's $r$),
the Spearman rank-order correlation coefficient, and Kendall's $\tau$
\citep{press1992}, we test for trends in $\delta r_i$ and $\delta z_i$
with time. We then compute the standard deviation of $\delta r_i$
($\sigma_{r,i}$) and $\delta z_i$ ($\sigma_{z,i}$). As a second test,
we divide the snapshots into groups of 100, derive $r_{avg,i}$, $\delta r_i$,
and $\sigma_{r,i}$ (and the corresponding $z$ variables) within each group,
and search for trends of these variables from the first group of 100 snapshots 
to the last group of 100 snapshots.

Relative to a system where the four small satellites have zero mass, we
consider whether an apparently stable light system has a large
$\sigma_{r,i}$ ($\sigma_{z,i}$) or a significant trend of $\delta r_i$
($\delta z_i$) with time. In a stable system, the typical dispersion in
$\delta r_i$ is small, ranging from 0.01~$R_\oplus$ for Nix to 0.02~$R_\oplus$
for Hydra. There is no measurable trend of $\delta r_i$ with time: the 
three correlation coefficients are indistinguishable from zero at a high 
confidence level (probabilities $p > 0.3$). The dispersions in $\delta z_i$ 
are a factor of 2--4 smaller than those in $\delta r_i$, with similarly 
small evidence of a trend with time, $p > 0.5$. The lack of trends in
the eccentricities (as measured by $\delta r_i$) or the inclinations
(as measured by $\delta z_i$) with time suggest the orbits are stable.

In very unstable systems, the dispersion in $\delta r_i$ for Styx and
Kerberos is somewhat larger and the three correlation coefficients are
positive with probabilities $p < 10^{-10}$ that the coefficients are
consistent with zero. Often, there are also clear trends in $\delta z_i$
with time for Styx and Kerberos, with equally low probabilities that the 
correlation coefficients are consistent with zero.

There are a few systems where the trends in $\delta r_i$ and $\delta z_i$
with time are less obvious.  Here, we rely on the Pearson, Spearman, and 
Kendall tests.  When all probabilities from these tests are small, 
$p < 10^{-3}$, the trend with time is significant at the 3-$\sigma$ 
(or better) level.  We judge a system unstable.  Longer-term integrations 
would likely yield more significant trends in $\delta r_i$ with time for 
these systems. When $10^{-2} > p > 10^{-3}$, trends of $\delta r_i$ or
$\delta z_i$ with time are not significant. These systems are marginally 
stable.

\vskip 4ex
\subsection{Main Results}
\label{sec: results}

\begin{figure}
\begin{center}
\includegraphics[width=5.5in]{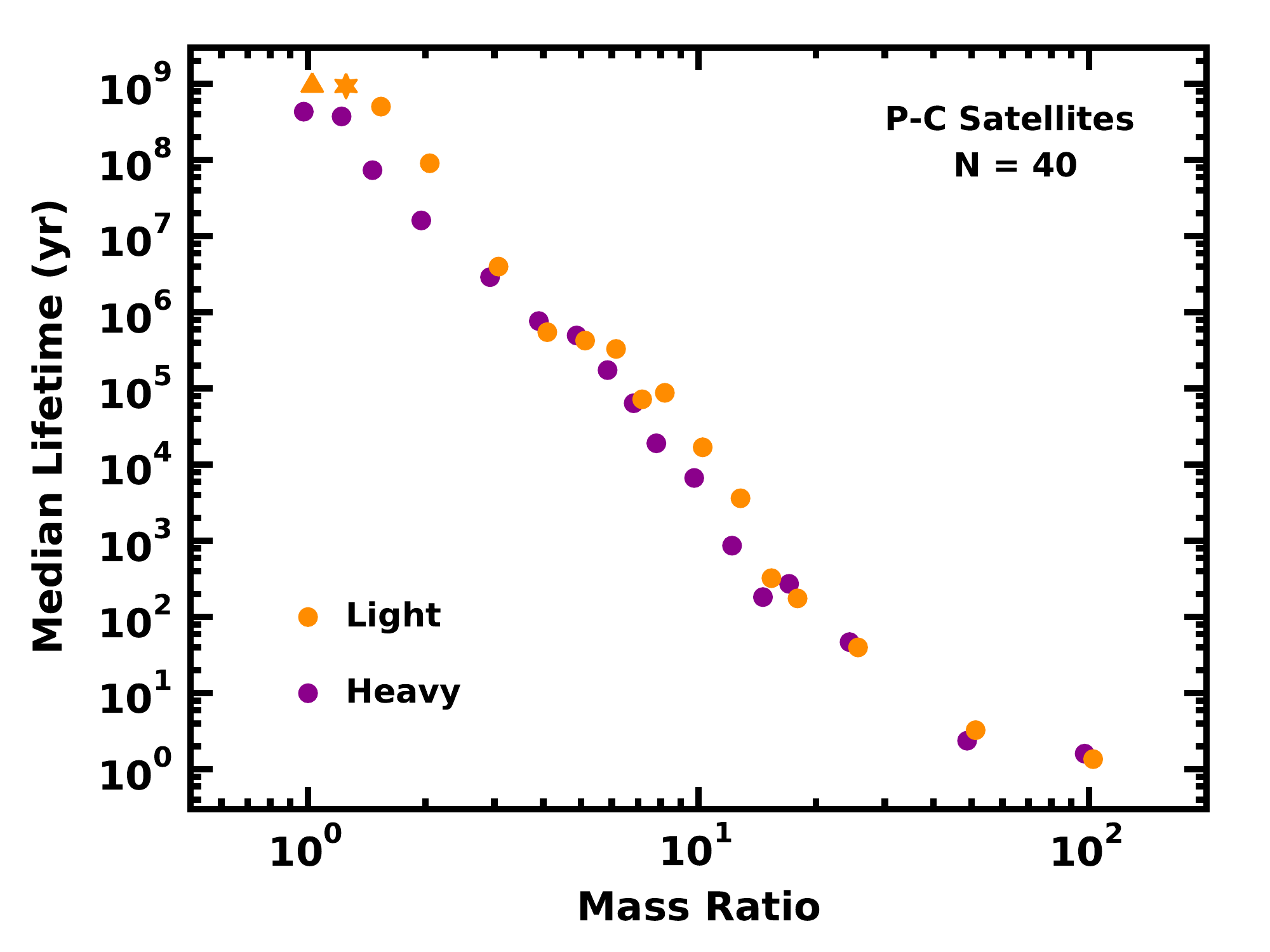}
\end{center}
\vskip -5ex
\caption{
\label{fig: all40}
  Median lifetime $\tau$ as a function of mass ratio $f$ for
  light (orange symbols) and heavy (purple symbols) satellite systems.
  Calculations yield lower limits on $\tau$ for marginally unstable light
  systems with $f$ = 1.0 (orange triangle) and clear unstable light 
  systems with $f$ = 1.25 (orange hexagon).
}
\end{figure}

In calculations where we multiply the masses of the small satellites by
the same factor, $f$, the lifetime is very sensitive to the total mass
(Fig.~\ref{fig: all40}). Among the 7--10 calculations of systems with
$f$ = 50 or $f$ = 100, at least one satellite is ejected within 1--3~yr.
As we decrease $f$, there is a clear progression in the median lifetime
$\tau_m$, from 10--100~yr for $f$ = 15--25 to 10~Myr for $f$ = 2.  Among
the calculations with $f \approx$ 1, lifetimes are 0.1--1~Gyr.

Heavy systems with $f \ge$ 1 are unstable.  For each $f \ge$ 1.25, all 
simulations eject at least one satellite.  Among the 14 simulations of 
heavy systems with $f$ = 1, eleven produce an ejection on time scales 
ranging from 70~Myr to 960~Myr. In two systems with no ejections, the 
orbital $e$ of Styx and Kerberos grows steadily with time. Only one of 
the 14 calculations maintains a nearly steady $e$ for Styx and Kerberos. 
Thus, there is a formal 93\% likelihood that a heavy satellite system 
with $f$ = 1 is unstable on time scales at least a factor of five smaller 
than the age of the solar system.

Light systems with $f > 1$ are also unstable.  At large $f \ge$ 3,
outcomes are chaotic; there is little difference in the lifetimes
of light and heavy systems.  When $f \le$ 2, lifetimes for light
satellite systems are 2--4 times longer than lifetimes for heavy
systems.  Among the configurations with $f \approx$ 1, none produce
an ejection after nearly 1~Gyr of dynamical evolution. A few, however,
show evidence for a slowly increasing $e$ in Styx or Kerberos or both.
Thus, light systems with $f \approx$ 1 are marginally unstable. on 1~Gyr
time scales

\begin{figure}
\begin{center}
\includegraphics[width=5.5in]{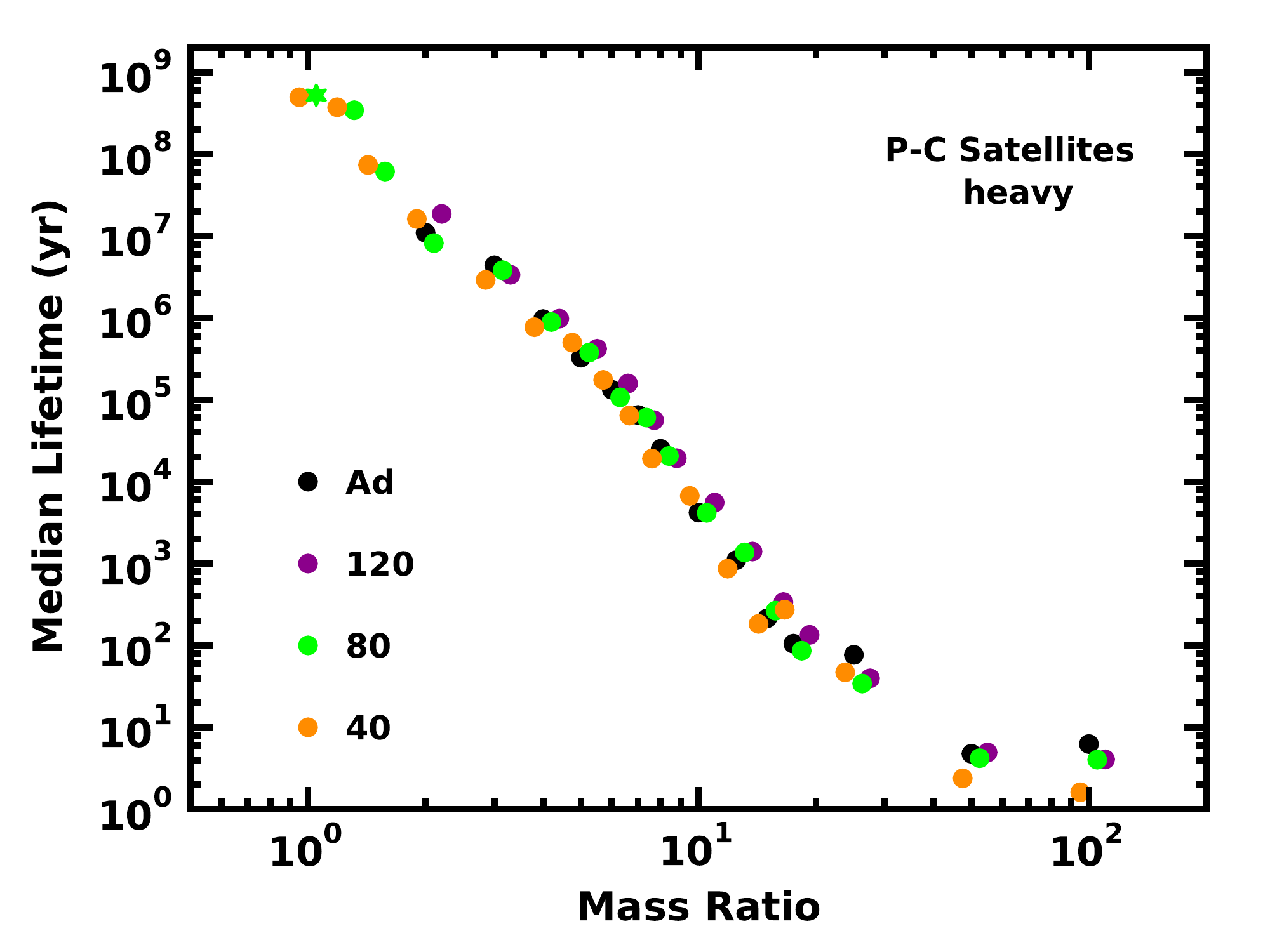}
\end{center}
\vskip -5ex
\caption{
\label{fig: int-comp}
Median lifetime $\tau$ for heavy satellite systems as a function of mass
ratio $f$ for four different integration routines as listed in the legend.
Calculations with $N$ = 80 yield lower limits on $\tau$ for systems with
the nominal masses, $f$ = 1 (starred symbols).  Orbits of systems with
initial masses $f$ times larger than the nominal satellite masses become
unstable on time scales ranging from 1--10~yr ($f$ = 50--100) to
$10^7 - 10^8$~yr ($f$ = 1.5--2).
}
\end{figure}

These results are independent of the integrator (Fig.~\ref{fig: int-comp}).
For heavy systems with $f$ = 2--100, calculations with $N$ = 40, 80, and
120 yield the same $\tau_m$. The extra time spent to resolve close encounters
with the adaptive integrator also has little impact on $\tau_m$ for $f$ = 2--100.
When $f <$ 2, symplectic integrations with $N \ge$ 100 or adaptive integrations
are too computationally intensive. However, symplectic integrations with 
$N$ = 80 yield the same median lifetimes for $f$ = 1 and $f$ = 1.5 as those 
with $N$ = 40.

Results for light systems are similar. For these calculations, we added an
additional comparison with $N$ = 150. As with the heavy systems, the median
lifetimes for $f \ge$ 2 are independent of the method of integration. For
$f$ = 1.25 and 1.5, symplectic integrations with $N$ = 80 yield similar $\tau_m$
as those with $N$ = 40.

\begin{figure}
\begin{center}
\includegraphics[width=5.5in]{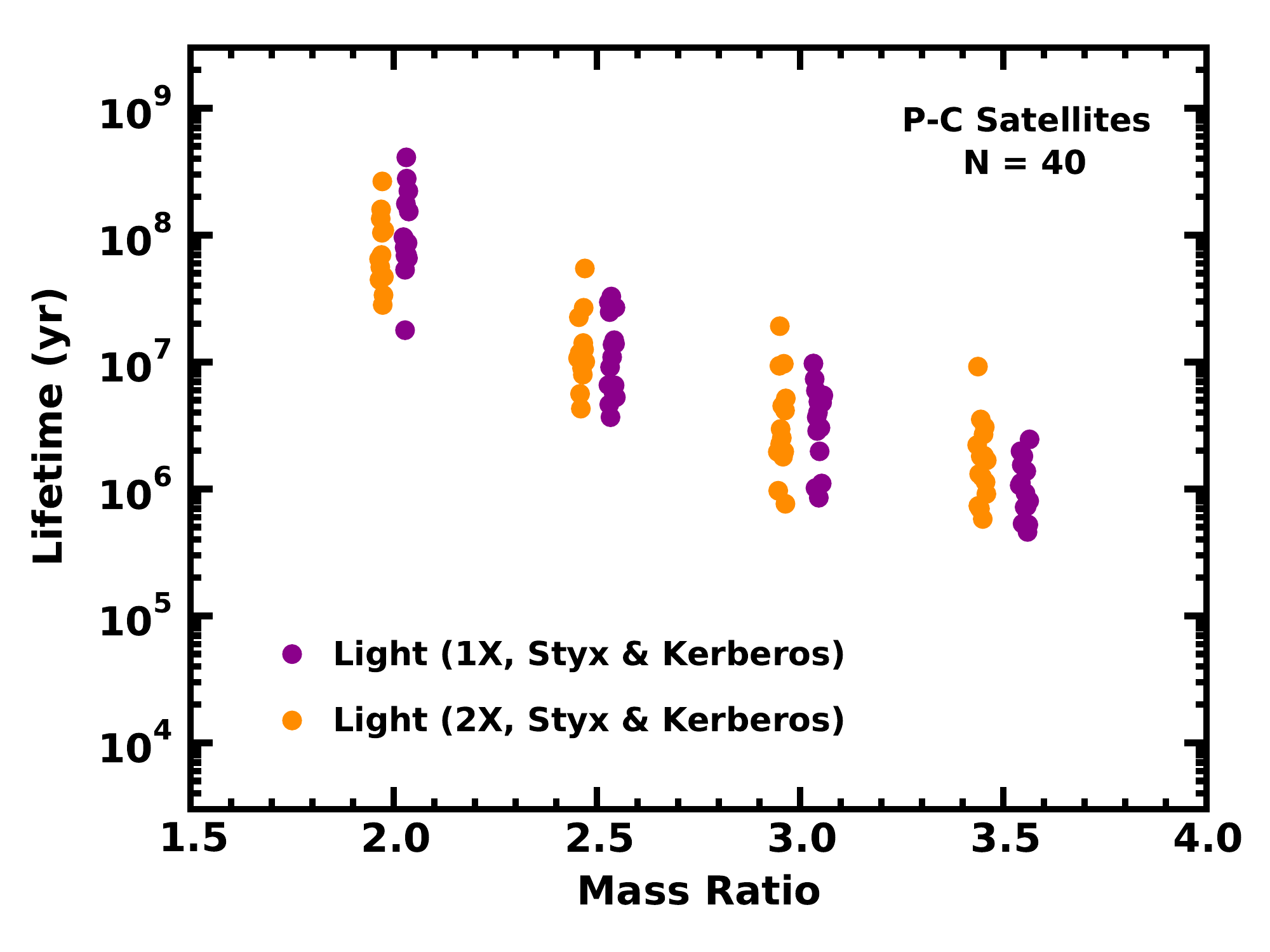}
\end{center}
\vskip -5ex
\caption{
\label{fig: light2x}
Comparison of lifetimes $\tau$ for the complete ensemble of calculations
with $f$ = 2--3.5 for light satellite systems with the nominal masses for
all satellites (purple symbols) and with double the nominal masses for
Styx and Kerberos (orange symbols).
}
\end{figure}

To check the sensitivity to input parameters in more detail, we
consider calculations where Styx and Kerberos have twice their
nominal masses in a light satellite system (Fig.~\ref{fig: light2x}).
Results for $f$ = 2.5--3.5 show similar lifetimes for the two
different masses of Styx and Kerberos. When $f$ = 2, the median
lifetime for a light system with more massive Styx and Kerberos,
$\tau \approx$ 60~Myr, is shorter than the median, $\tau \approx$
90~Myr, for a light system with the nominal masses for Styx and
Kerberos. Both median lifetimes are much shorter than the 4.5~Gyr
age of the solar system. However, a KS test returns a probability
of 15\% that the two sets of lifetimes are drawn from the same
parent distribution. Thus, the distributions of lifetimes are
formally indistinguishable.

These results confirm the expectations of dynamical theory.  Heavy systems
with $f \ge$ 1 are unstable on time scales much shorter than the age of the
solar system. Although they are more stable than heavy systems, light systems
with $f \ge$ 1.25 are also unstable on relatively short time scales. Light
systems with $f \approx$ 1 are marginally unstable on a 1~Gyr time scale. 
Although we have not completed calculations for light systems with twice 
the nominal mass of Styx and Kerberos and $f \le$ 1.5, results for $f$ = 
2--3.5 suggest the lifetimes are fairly independent of the masses of Styx 
and Kerberos.

To conclude this section, we examine sets of simulations where we augment
the mass of one satellite ($f_i >$ 1 for $i$ either N, K, or H) and keep
the masses of the other satellites at their nominal HST masses. We keep
the mass of Styx fixed; its small mass precludes much improvement with
$n$-body models.  Instead of performing multiple calculations with very
similar $f_i$, we derive lifetimes $\tau$ for calculations that densely
sample $f_i = $ 1--6.

\begin{figure}
\begin{center}
\includegraphics[width=5.5in]{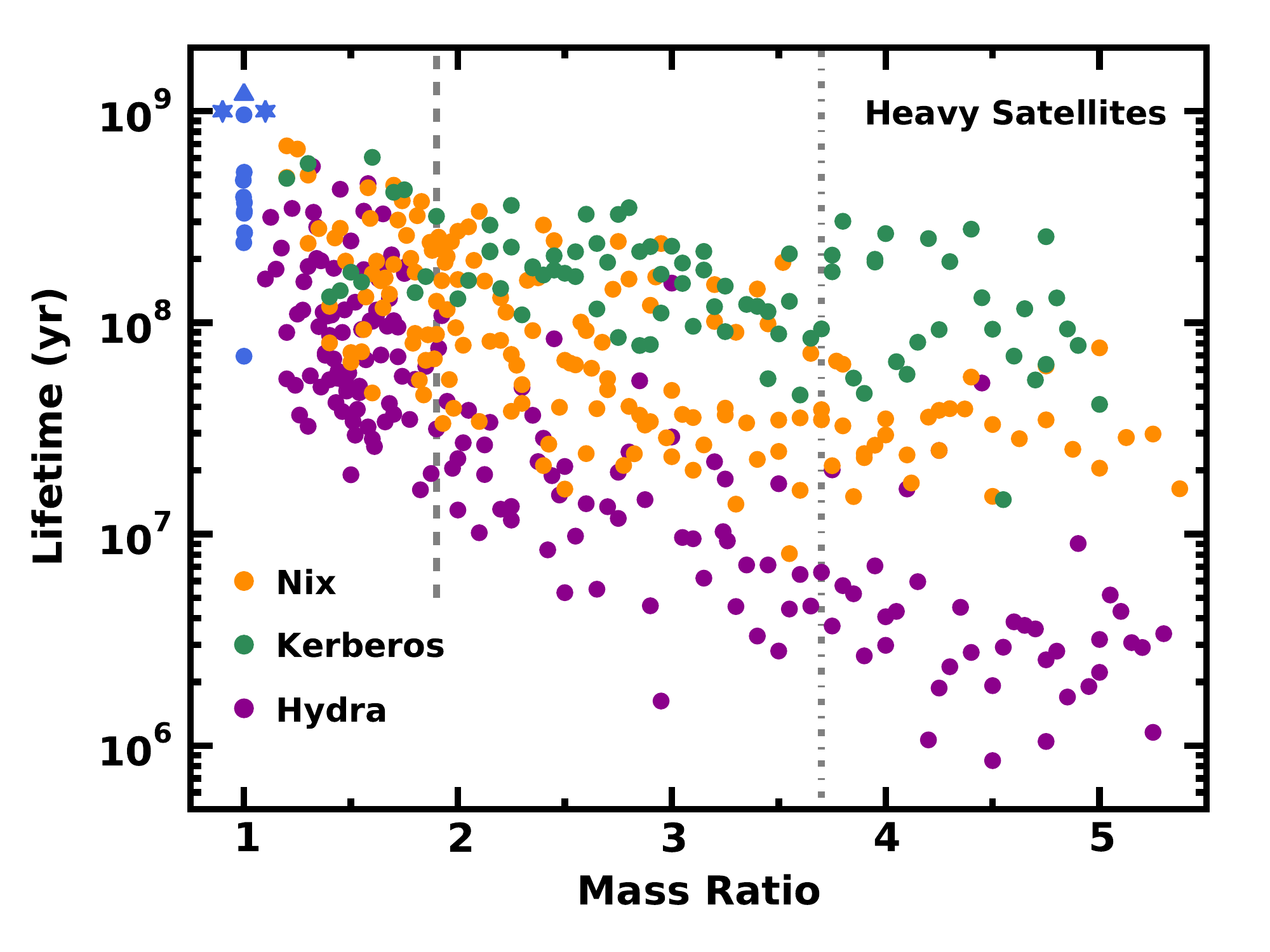}
\end{center}
\label{fig: f-nkh}
\vskip -5ex
\caption{
Lifetimes for heavy satellite systems with $f_i > 1$ for Nix
  (orange points), Kerberos (green points), or Hydra (purple points)
  and $f$ = 1 for the other satellites.  Several points have been
  displaced horizontally or vertically by small amounts for clarity.
  The vertical dashed (dotted) lines indicate 1$\sigma$ (3$\sigma$)
  upper limits on the masses from fits to HST data.  Among the 14
  distinct calculations with $f$ = 1 for all satellites (blue symbols),
  eleven resulted in at least one ejected satellite (filled circles),
  two are unstable with slowly growing $e$ for Styx and Kerberos but
  no ejection (filled hexagons), and one appears stable with an
  oscillating $e$ for Styx and Kerberos (filled triangle).
}
\end{figure}

The results confirm that heavy satellite systems are unstable
(Fig.~\ref{fig: f-nkh}).  Models with $f_H$ = 1--1.4 have $\tau \approx$
50--600~Myr.  Although several calculations with larger mass ratios
have similarly long lifetimes, $\tau$ declines monotonically with
$f_H$, reaching $\tau \approx$ a few Myr for $f_H$ = 4--5.
Calculations with a more massive Nix show a shallower variation of
lifetime with $f_N$, ranging from $\tau \approx$ 20--40~Myr for $f_N
\approx$ 4--5 to 60--700~Myr for $f_N \approx$ 1--1.6. Following this
trend, the system lifetime has an even shallower dependence on the
mass of Kerberos, with $\tau \approx$ 50--600~Myr for $f_K$ = 1--5.

In all of these simulations, systems where the mass of one satellite is
larger than the nominal masses have much longer lifetimes than systems
where all of the satellites are more massive. As an example, heavy
systems with $f$ = 4--5 for all satellites have median lifetimes
$ \tau_m \sim$ 1~Myr. Systems with $f_H$ = 4--5 are somewhat more stable,
with lifetimes of 1--10~Myr. When $f_N$ = 4--5, lifetimes are much longer,
10--100~Myr. Calculations with $f_K$ = 4--5 are even more stable, with
lifetimes of 50--300~Myr.

This behavior correlates with the nominal masses of the three satellites.
As the outermost and most massive satellite, Hydra has a significant impact
on the dynamical evolution of the inner satellites. Making Hydra more massive
tends to push the inner satellites towards the \pc\ binary. As \pc\ pushes
back, the satellite system becomes unstable. In contrast, a more massive Nix
or Kerberos tends to push Hydra away from the inner binary. With less pushback
from \pc, the more massive satellite system can then occupy a somewhat larger
volume than the nominal system and have a somewhat longer lifetime.

\begin{figure}
\begin{center}
\includegraphics[width=5.5in]{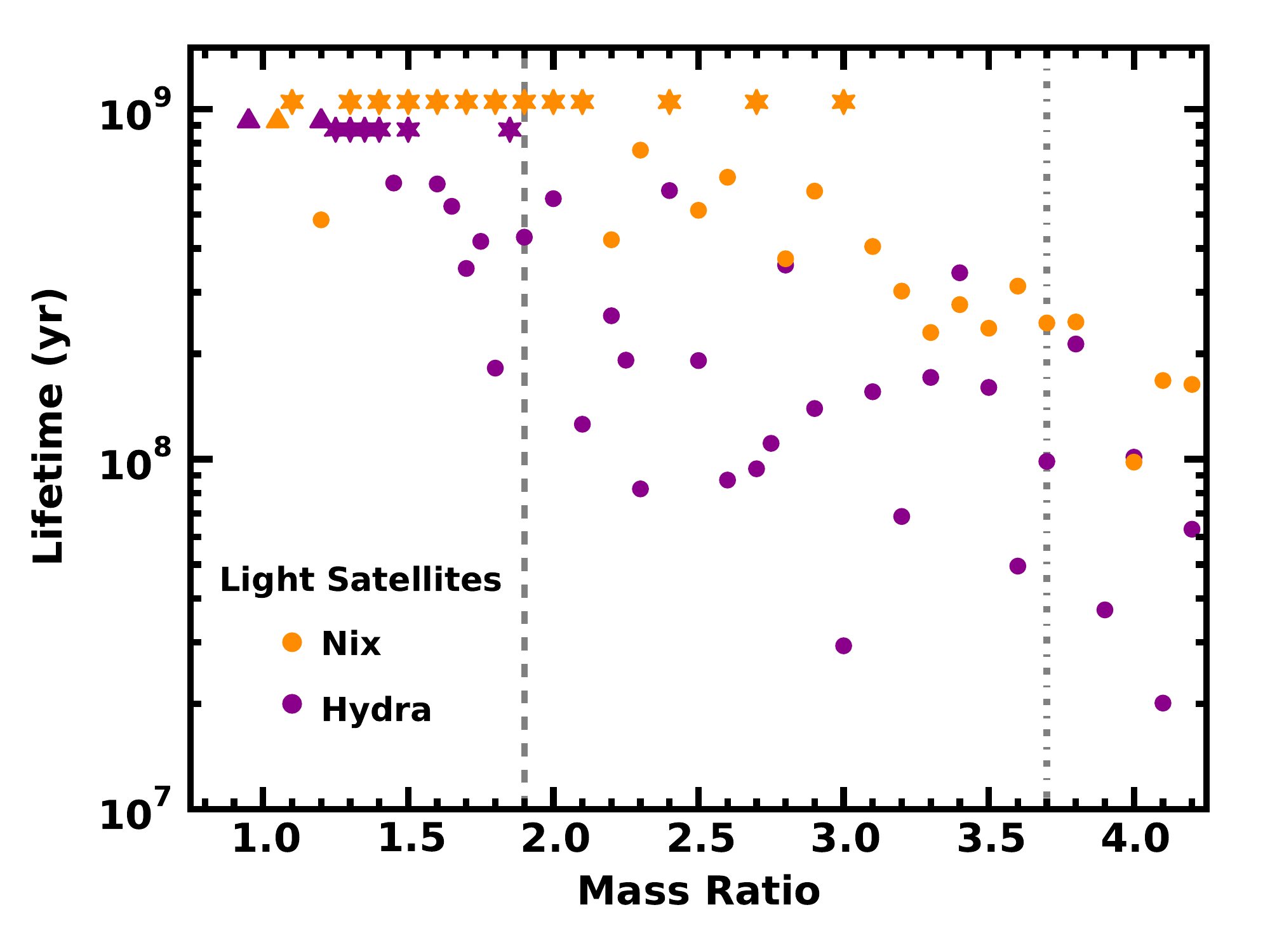}
\end{center}
\vskip -5ex
\label{fig: f-nh}
\caption{
Lifetime $\tau$ for light satellite systems where the mass of Nix
(orange symbols) or Hydra (purple symbols) has been augmented by a
factor $f$ = 1--5. Hexagons (triangles) indicate lower limits on the
lifetime for unstable (possibly stable) systems. Some points have been
displaced vertically by 0.1 in the log for clarity. Gray vertical dashed
(dot-dashed) lines mark 1$\sigma$ (3$\sigma$) limits on masses from HST
data. Light systems with $f_N \ge$ 1.1 or $f_H \ge$ 1.15 are unstable.
}
\end{figure}

Repeating this exercise with the light satellite system leads to similar
conclusions (Fig.~\ref{fig: f-nh}). Compared to the heavy system, light
systems with $f_H$ = 3--4 survive $\sim$ ten times longer before ejecting
either Styx or Kerberos.  For $f_H$ = 3--4, the typical lifetime is 20--200~Myr.
Systems with smaller $f_H$ have significantly longer lifetimes, with
$\tau \approx$ 100--500~Myr at $f_H$ = 2 and $\tau \ge$ 400~Myr at $f_H \le$ 1.75.
Among the calculations with $f_H \le$ 1.75, several produce an ejected satellite.
In others, the orbital $e$ of Styx and Kerberos steadily increase in time.
Although a light system with $f_H$ = 1.2 shows few signs of instability after
975~Myr of dynamical evolution, another system with $f_H$ = 1.1 is clearly
unstable.

Systems with $f_N > 1$ generally last much longer than those with $f_H > 1$.
Typical lifetimes range from $\sim$ 100~Myr at $f_N$ = 4 to $\sim$ 400~Myr
at $f_N$ = 2.5--3.5 to more than 1~Gyr at $f_N$ = 1. Curiously, all of the
calculations with $f_N \ge$ 1.1 show a steadily increasing $e$ with time for
Styx and Kerberos. On times scales of 1--2~Gyr, we expect each of these
calculations will result in an ejection of Styx or Kerberos.

\vskip 4ex
\subsection{Summary: Robust Satellite Masses}

The \nbody\ calculations suggest that the heavy (light) satellite system
with the nominal masses is clearly (probably) unstable.  Calculations for
the light satellite system place the strongest constraints on the masses 
of Nix and Hydra.  After $\sim$ 1~Gyr for many of 14 orbital integrations, 
the light system with $f$ = 1 occupies an unstable state where the $e$ of 
Styx and Kerberos gradually increase with time.  Systems with $f \ge$ 1.25 
eject at least one small satellite on time scales ranging from 1--10~yr 
($f$ = 50--100) to $10^4$~yr ($f$ = 10) to 100--300~Myr ($f$ = 1.25--1.50).  

Calculations where we vary the masses of Hydra or Nix independently of the 
other satellites yield robust upper limits for stable light systems:
$f_H \le$ 1.15 (when $f_N$ = 1) and $f_N < $ 1.1 (when $f_H$ = 1).  Adopting 
$f \approx f_N \approx f_H$ implies an upper limit $f \le$ 1.05 for stability.
Despite the smaller number of these calculations, this upper limit agrees
reasonably well with the limit $f \le$ 1 derived from calculations of light
systems with the nominal masses. 

These calculations follows a long tradition of using stability to constrain 
the masses and orbital elements of known circumstellar and circumbinary 
planet and satellite systems \citep[e.g.,][and references therein]{duncan1997,
ito2001,fabrycky2010,french2012,youdin2012,mahajan2014,obertas2017}.  In 
many of these studies, stability is inferred from fits to a set of direct 
\nbody\ calculations \citep[e.g.,][]{duncan1997,french2012,youdin2012}.  The 
\nbody\ results in these examples yield a relation between the lifetime of 
the system $t$ and the mass factor $f$,
\begin{equation}
t = t_0 f^{-\beta} ~ .
\label{eq: t-power}
\end{equation}
When the processor time required to calculate stability for some range of
$f$ is prohibitive, results for large $f$ are extrapolated to small $f$.

Our calculations provide a way to test this approach for the \pc\ system. 
Fits to the full ensemble of calculations for $f$ = 2--100 yield $t_0$ 
and $\beta$. For the heavy satellite system, we may then compare the 
predicted $t$ at $f$ = 1 with the time scales inferred directly.
For the light system, the \nbody\ calculations provide good evidence for
instability on time scales of $\sim$ 2~Gyr. Comparing this time scale 
with the predicted $t$ places stronger constraints on satellite masses.

To perform these fits, we employ the robust estimation routine MEDFIT 
\citep{press1992}, which fits a straight line to a set of points by 
minimizing the absolute deviation.  Converting Eq.~\ref{eq: t-power} to 
a linear equation, fits to the data for the heavy system for $N$ = 40 
and $f$ = 1--100 result in $t_0$ = 550~Myr and $\beta \approx$ 4.9. 
Considering the range $f$ = 2--100 yields very similar fits for $N$ = 
40, 80, and 120 and for the adaptive integrator, $t_0$ = 300--600~Myr 
and $\beta$ = 4.8. Contracting the range in $f$ to 3--100 or 4--100 has
little impact: $t_0$ = 600--700~Myr and $\beta$ = 5 for $f$ = 3 and
$t_0$ = 500--700~Myr and $\beta$ = 4.9 for $f$ = 4. In all of these fits,
the absolute deviation of the points from the fit is 0.3--0.4.  Analyzing 
only those results with $f$ = 5--100 or $f$ = 6--100, however, degrades
the quality of the fit; the absolute deviation and range in $t_0$ are 
then much larger.  

For comparison, \citet{youdin2012} derive $t_0 \ge$ 200~Myr and $\beta$ = 
3.6--4.6 for calculations of \pc, Hydra, Kerberos, and Nix with $f \ge$ 4--10. 
From the full suite of \nbody\ calculations described here, the median 
lifetime of a heavy system with $f$ = 1 is 430~Myr with a full range of 
70~Myr to 960~Myr. Fits to results for $f$ = 2--100 capture this range 
rather well, predicting an average $t_0$ = 510~Myr. Adopting the absolute 
deviation as a measure of the full range of $t$ at any $f$ suggests a 
minimum $t \approx $ 250~Myr and a maximum $t \approx$ 1000~Myr, close 
to the range derived from the \nbody\ calculations.

Repeating this analysis for the \nbody\ calculations where we multiply the
mass of only one satellite by a factor $f > 1$ yields similar results. Fits
to the \nbody\ data for a heavy satellite system with $f_H \ge$ 1.5 yield 
$t_0$ = 300~Myr and $\beta$ = 3.  Similar data for $f_N \ge$ 1.5 ($f_K \ge$ 
1.5) generate $t_0$ = 375~Myr and $\beta$ = 1.75 ($t_0$ = 410~Myr and 
$\beta$ = 0.925). Taken together, the $t_0$ implied by this suite of 
\nbody\ calculations for the heavy system agrees with the $t_0$ derived 
from those with a common $f$ for all satellites.

For the light system, the fit to the full ensemble of results for $N$ = 40
and $f$ = 1--100 returns $t_0 \approx$ 2400~Myr and $\beta \approx$ 5.3.
Removing data for $f <$ 2 allows a comparison for calculations with different
$N$ and with the adaptive integrator: $t_0$ = 1500--2200~Myr and $\beta$ =
5.1--5.3. There is also little difference among the various fits for 
$f$ = 3--100, $f$ = 4--100, and $f$ = 5--100; all of the integrators suggest 
$t_0$ = 1000--3000~Myr and $\beta$ = 5.0--5.3. Removing more data from the
analysis leads to a much larger range in $t_0$ and $\beta$ and generally 
larger absolute deviations.  Although we cannot make a direct comparison 
between the lifetimes derived from \nbody\ calculations and these fits,
the growth in $e$ and $\imath$ of light systems with $f$ = 1 suggests 
lifetimes $\sim$ 2000~Myr. The two analyses clearly agree.

Among the \nbody\ calculations for light systems with $f_H \neq 1 $ or 
$f_N \neq 1$, only the data for $f_H \neq 1$ cover a sufficiently large 
range in $f$ to perform a high quality fit. With $t_0$ = 1600~Myr and 
$\beta$ = 2.6, the lifetime derived for these simulations is identical 
to the 2000~Myr lifetime implied by calculations with the same $f$ for
all satellites.

Overall, this examination demonstrates a set of robust upper limits for 
the masses of Hydra and Nix. From the fits to the \nbody\ simulations
with identical $f$ for all satellites, light systems with $f$ = 0.75, 
0.85, and 1.0 have median lifetimes of 10~Gyr, 4.5~Gyr, and 2~Gyr. Based 
on the factor of 2--3 dispersion in lifetimes among any calculation with 
fixed $f$, we expect survival rates of $\sim$ 0\% ($f$ = 1.05-1.10), 
10\%--20\% ($f$ = 1.0), 50\% ($f$ = 0.85), or 90\%--100\% ($f$ = 0.75) 
for the 4.5~Gyr age of the solar system. Thus, reasonable upper limits 
to the masses of Nix and Hydra are $\sim$ 10\% larger than their nominal 
masses. Among the calculations with $f_H > 1$ or $f_N >$ 1, systems with 
$f_H$ somewhat larger than 1 are more stable than those with $f_N$ somewhat
larger than 1. We therefore set the upper limits $f_N \le$ 1.05 and 
$f_H \le$ 1.15; together, these correspond well with $f \le$ 1.1.

The \nbody\ calculations in this study place weaker limits on the masses
of Styx and Kerberos. Requiring at least one heavy system survive for 
4.5~Gyr implies $t$ = 1500~Myr, $f \le$ 0.8, and a total mass for the
four satellites $\sim 9.1 \times 10^{19}$~g. This mass is comparable to 
the mass of a light system with $f \approx$ 1, $\sim 9.3 \times 10^{19}$~g. 
Thus, the \nbody\ results for the heavy systems are consistent with the 
small mass for Kerberos and Styx adopted for the light systems. Once we 
complete calculations for light systems with $f$ = 1.5 and 2 and twice 
the nominal masses for Styx and Kerberos, it should be possible to place 
better limits on the masses of Styx and Kerberos.

\vskip 6ex
\section{SATELLITE MASS DENSITY}
\label{sec: density}

To estimate the mass density of Nix and Hydra, we consider the volume $V$ within
three types of solids: boxes, ellipsoids, and pyramids. Defining $a$, $b$, $c$
as the lengths of the semiaxes with $a > b > c$, $V$ = $8 a b c$ (box),
$4 \pi a b c / 3$ (ellipsoid), or $8 a b c / 3$ (pyramid). Of these options,
approximating the satellites as boxes (pyramids) yields the largest (smallest)
volume and the smallest (largest) mass density. As a plausible compromise, we
derive the mass density from the volume of an ellipsoid. We infer
$\rho_N \le$ 1.57~g~cm$^{-3}$ as the mass density for Nix ($f_N \le$ 1.05)
and $\rho_H \le$ 1.44~g~cm$^{-3}$ for Hydra ($f_H \le$ 1.15).

If all satellites have the same mass density, we can estimate $\rho$ from the
total mass, $m \lesssim 10^{19}$~g, established in the \nbody\ and the total
volume from the \nh\ size measurements in Table~\ref{tab: sats}.  The result,
$\rho \lesssim$ 1.3~\gcmc, is somewhat smaller than the upper limits on the
mass density soley from Nix and Hydra. 

Although these mass densities are smaller than the mass density for Charon, they
do not include the uncertainties in the measured sizes. Rather than estimate a
plausible range in an upper limit for the mass density from the errors in sizes,
we estimate the probability of a particular mass density from a Monte Carlo
calculation. The calculation assigns random sizes
\begin{equation}
x_i = x_{i, 0} + f_g \delta x_i ~ ,
\label{eq: mcsize}
\end{equation}
to each satellite, where $x$ is either $a$, $b$, or $c$, $i$ is N (for Nix) or
H (for Hydra), and $f_g$ is a gaussian deviate from a random number generator.
The subscript `0' refers to the measured length of the semiaxis. With $a$, $b$,
and $c$ known, the volume $V_i$ for a box, ellipsoid or pyramid follows.

To choose the mass, we consider three approaches. In the simplest estimate,
we adopt the upper limit derived from the $n$-body calculations,
\begin{equation}
m_i = m_{u,i} ,
\label{eq: mcmass1}
\end{equation}
where $m_{u,i}$ is 5\% (15\%) larger than the nominal mass for Nix (Hydra).
As a second approach, we consider the nominal mass $m_{i,0}$ and adopted
error $\delta m_i$ from the $n$-body calculations, deriving a model mass
\begin{equation}
m_i = m_{i, 0} + f_g \delta m_i ~ .
\label{eq: mcmass2}
\end{equation}
The mass density is then $\rho_i = m_i / V_i$. Finally , we adopt a lower limit
to the mass $m_{i,l}$ and set the mass as
\begin{equation}
m_i = m_{l,i} + f (m_{u,i} - m_{l,i}) ~ ,
\label{eq: mcmass3}
\end{equation}
where $f$ is a uniform deviate between 0 and 1.
Repeating each procedure $N$ times yields three probability distributions for
$\rho_i$.

\begin{figure}
\begin{center}
\includegraphics[width=5.5in]{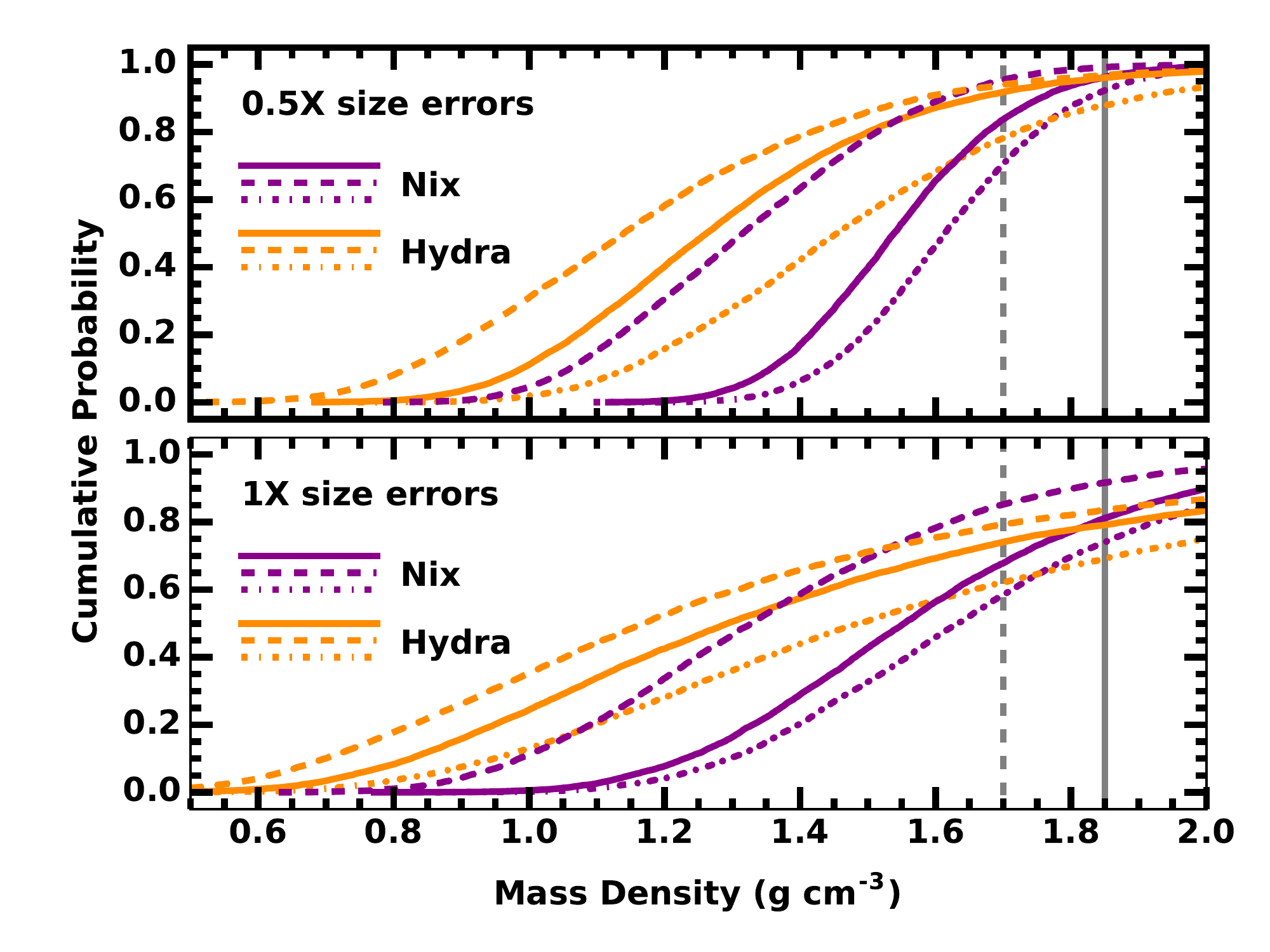}
\end{center}
\vskip -5ex
\caption{
\label{fig: rho}
Cumulative probability that Nix (purple curves) and Hydra
(orange curves) have a density smaller than $\rho$ using mass model 1
(dot-dashed curves), mass model 2 (solid curves), or mass model 3
(dashed curves) for the Monte Carlo calculations described in the text.
Lower panel: volume derived from the nominal errors.
Upper panel: volume derived from errors half the size of the nominal
errors. Vertical dashed (solid) grey lines indicate the measured mass
density of Charon (Pluto). For plausible masses of Nix and Hydra,
their mass density is most likely smaller than the mass density of
Charon and significantly smaller than the mass density of Pluto.
}
\end{figure}

Fig.~\ref{fig: rho} and Table~\ref{tab: rho} summarize the results.
The cumulative probability distributions
$p(\rho < \rho)$ have several characteristic features. Due to the smaller errors in
its size, Nix has sharper distributions for all mass models; $p(\rho)$ for Hydra is
much broader.  Given the smaller volume of Nix relative to Hydra, the probability that
Nix has a mass density smaller than the mass density of Charon is smaller than the
corresponding probability for Hydra for any mass model. At larger mass densities, this
behavior reverses: for two of three mass models, Nix has a higher probability of having
mass density smaller than the mass density of Pluto.

These results suggest that the mass densities of Nix and Hydra are smaller than those
of Charon and Pluto. In the most conservative mass model 1, roughly 59\% (62\%) of the
Monte Carlo trials yield a mass density for Nix (Hydra) smaller than that of Charon
(see Table~\ref{tab: rho}). In the most liberal model 3, these probabilities grow to
$\sim$ 80\%.
From our calculations, the most likely masses for Nix and Hydra are close to their
nominal masses (e.g., model 2). In this picture, Nix and Hydra have mass densities
smaller than Charon $\sim$ 70\% of the time.

The Monte Carlo calculations demonstrate that reducing errors in the size estimates for
Nix and Hydra place much better constraints on the mass density. For the nominal sizes,
reducing the errors by a factor of two improves the likelihood of mass densities smaller
than Charon by as much as 15\% to 20\%. Stronger limits on the masses generate weaker
improvements of the probabilities.

Reducing the uncertainties in the satellite shapes would also enable better estimates
for the mass density.  From published {\it New Horizons} images, Nix does not resemble
a pyramid and looks more like an ellipsoid than a box.  Deriving a 3D shape from the
full ensemble of {\it New Horizons} images would eliminate this ambiguity.  For Hydra,
the poorer image resolution complicates volume estimates. As with Nix, using the complete
set of imaging data would considerably reduce uncertainties in the volume.

\vskip 6ex
\section{DISCUSSION}
\label{sec: disc}

The new limits on the masses for the \pc\ satellites allow us to
(i) compare their properties with other satellites in the solar system,
(ii) improve our understanding of the stability of circumbinary 
planetary/satellite systems, and
(iii) examine how the satellites find their places after the giant 
impact that formed the central binary. Before summarizing our overall
conclusions, we briefly consider each of these topics.

\vskip 6ex
\subsection{Comparison with Other Satellite Systems}

Starting with Mars (Phobos and Deimos) and continuing outward
from the Sun with Jupiter 
(Carme, Metis, and Sinope), Saturn (Atlas, Daphnis, Helen, and Pan), 
Uranus (Cordelia and Ophelia), and Neptune (Laomedeia, Psamanthe, 
and Sao), satellites with radii of 5--20~km are common throughout
the solar system \citep[e.g.,][]{thomas1989,thomas1998,kark2001,
rettig2001,kark2003,thomas2010,sheppard2006,slyuta2014}. Imaging 
data from \voy, \cas, \nh, and various missions to Mars reveal a 
variety of shapes and structures on the surfaces of these small
satellites.  

Despite extensive knowledge about their surface characteristics, 
measurements of the mass density for small satellites (radii less
than 50--100~km) are rare.  The densities of Phobos (1.9~\gcmc) 
and Deimos (1.5~\gcmc) are much larger than those of the somewhat 
larger Prometheus (0.5~\gcmc) and Pandora 
\citep[0.5~g~cm$^{-3}$;][]{jacobson2004,renner2005,jacobson2010,
patz2014}. The density of Uranus' moon Cressida (0.9~\gcmc) is 
roughly midway between the two Martian satellite and Saturn's 
satellites orbiting close to or within the ring system 
\citep{chancia2017}.  Estimates for the densities of other small
satellites\footnote{https://ssd.jpl.nasa.gov/} rely on estimates 
instead of direct measurement of their masses.

Mass density estimates for Nix and Hydra place them within the broad 
range of measured densities of satellites with similar sizes. Their 
densities are clearly smaller than the density of Phobos and probably 
comparable to the density of Deimos.  Due to the lack of other satellites 
in the system, the factor of three reduction in mass required to have 
mass densities comparable to Daphnis, Prometheus, Pandora, and the 
other ring moons of Saturn seems unlikely \citep{kb2019}. Discovery of 
other small satellites with orbits between Styx and Hydra would test 
this assertion.

Pluto's small satellites distinguish themselves with their large albedos, 
ranging from 0.56 for Styx and Kerberos to 0.83 for Hydra \citep{weaver2016}.  
Despite their similar mass density, albedos of Phobos and Deimos, $\approx$ 
0.07, are roughly an order of magnitude smaller than the albedos of Styx, 
Nix, Kerberos, and Hydra \citep{zellner1974, thomas1996,cantor1999}. Aside 
from Triton \citep[0.72;][]{hicks2004}, all of the satellites of Uranus 
and Neptune have albedos smaller than 0.4 \citep[e.g.,][and references 
therein]{kark2001,kark2003,fry2007,farkas2017}.  Among Jupiter's satellites, 
only Io, Europa, and Ganymede have albedos larger than 0.4 \citep{buratti1983,
simonelli1984}.  In contrast, many of Saturn's satellites have albedos 
comparable to those of Pluto's small satellites \citep{verbiscer2007,
pitman2010}.  Among the smaller satellites, Helene and Calypso have albedos 
of order unity \citep[see also][and references therein]{madden2018}. 

The similarity of the \pc\ and the small, inner satellites of Saturn 
could be a result of similar formation mechanisms. In the Saturn system, 
the origin of the rings remains controversial \citep[e.g.,][and references 
therein]{charnoz2009a,canup2010,hyodo2017,dubinski2019}. However, small
moons outside and within Saturn's rings likely grew from ring material and 
either migrated outside the rings \citep[e.g.,][]{charnoz2010} or generated 
gaps within the rings \citep[e.g.,][and references therein]{bk2013}.  The 
density of the moons and moonlets within the rings is then similar to the
low density of ring material. In the \pc\ system, the satellites form within
the icy debris of a giant impact, which has a lower density than Pluto or 
Charon, but probably has a larger density than the material in Saturn's rings. 

\vskip 6ex
\subsection{Circumbinary Dynamics}

As discussed in \S\ref{sec: theory}, there have been numerous studies 
into the stability of multi-planet (multi-satellite) systems orbiting 
single stars (planets) \citep[e.g.,][]{wisdom1980,petit1986,gladman1993,
chambers1996,deck2013,fang2013,kratter2014,fabrycky2014,mahajan2014,
pu2015,morrison2016,obertas2017,hwang2017,quarles2018a}. 
With two objects, stability requires a minimum separation in Hill 
units, $K = 2 \sqrt{3}$, that is independent of planet masses and 
orbital periods. In multi-planet systems, orbital resonances complicate
stability arguments. Although systems with three or more planets may 
be stable with $K \approx$ 6--7, systems with larger separations, 
$K \approx$ 10--12, are more generally stable for the age of a 
solar-type star.

With few studies of the stability circumbinary multi-planet systems
\citep[e.g.,][]{kratter2014,marzari2016}, results for the \pc\ system 
enhance our understanding of the circumbinary dynamics of multi-planet
systems. In our calculations, the heavy system with the nominal masses
has $K_{SN}$ = 12, $K_{NK}$ = 16, and $K_{KH}$ = 10; this system is 
clearly unstable on 500~Myr time scales. Fits to the full quite of 
simulations for heavy systems suggests a roughly 50\% survival rate 
on time scales of 10~Gyr for $f \approx$ 0.5, which is equivalent to 
a system with $K_{SN}$ = 15, $K_{NK}$ = 20, and $K_{KH}$ = 12.5.

Among light satellite systems, a 50\% survival rate over 10~Gyr requires
$f \approx$ 0.75. Satellite separations in Hill units are then nearly
identical to the separations of stable heavy systems, $K_{SN}$ = 14.5, 
$K_{NK}$ = 19, and $K_{KH}$ = 12. Generalizing to any satellite system
around \pc\ is challenging due to the orbital resonances \citep{smullen2017},
but it seems that this set of orbital separations would allow stability 
for other masses of the \pc\ satellites.

\vskip 6ex
\subsection{Formation Models}

The Pluto-Charon satellite system provides a fascinating challenge to planet formation
theories. Current ideas focus on the aftermath of a giant impact, where a glancing
collision between Pluto and Charon leads to an eccentric binary system with a period of
1--2~d \citep{mckinnon1989,canup2005,canup2011}.  Satellites grow in circumbinary debris
from the collision or in material captured afterwards \citep{stern2006,lith2008b,canup2011,
pires2012,kb2014b,walsh2015}.

This basic picture faces several hurdles. On a time scale of 1--10~Myr, tidal forces
circularize and expand the Pluto-Charon orbit \citep{farinella1979,dobro1997,peale2011,
cheng2014a}.  As the central binary evolves, orbital resonances pass through the volume
containing the debris \citep{ward2006,lith2008a,smullen2017}. These resonances pump the
eccentricities of circumbinary solids, destabilizing systems of satellites with properties
similar to those of the known satellites \citep{peale2011,cheng2014b,walsh2015,bk2015b,
smullen2017,woo2018}.  Although satellites embedded within rings of small particles survive
resonance pumping, the ensemble of small particles must be massive enough to damp the
orbits of larger satellites \citep{bk2015b}.

The time scale to grow satellites out of the debris is comparable to the circularization
time \citep{kb2014b,walsh2015}. In systems where the expansion of the binary is complete,
the number, masses, and orbital architecture of the satellites depend on the initial mass
of the debris:  more massive rings of debris favor fewer large satellites, while less
massive rings favor many small satellites. Numerical simulations match the current number
of satellites with an initial mass of $3-10 \times 10^{19}$~g for the debris \citep{kb2014b}.
The upper end of this range is consistent with the upper limit on satellite mass derived
here.  The \citet{kb2014b} simulations also predict several much smaller satellites with 
radii $R \le$ 1--3~km beyond the orbit of Hydra. Although new satellites have not been 
identified \citep{weaver2016}, a significant discovery space is accessible with the 
{\it James Webb Space Telescope} and other NASA missions \citep{kb2019,gaslac2019}.

Other aspects of the {\it New Horizons} data support this general model for satellite
formation. In a giant impact where Charon survives, the circumbinary debris should have
a larger proportion of ice than either Pluto or Charon. The densities derived for Nix 
and Hydra agree with this prediction.  The large measured albedos for all of the
satellites also imply an icier mixture than inferred for Pluto or Charon \citep{weaver2016}.
Finally, the shapes of the satellites also look like the products of an agglomeration
process \citep{weaver2016}.

The close packing of the satellites orbiting Pluto-Charon -- with orbital separations of
$K = 12-18$ in units of mutual Hill radii -- is reminiscent of several exoplanetary systems
discovered by the {\it Kepler} satellite -- including Kepler 11, Kepler 80, Kepler 90,
Kepler 223, and K2-138 -- where $n_p \ge$ 4--5 planets orbit near resonances and have close 
to the minimum orbital separation necessary for stability \citep{borucki2011,liss2011b,
mahajan2014,mills2016,macdonald2016,shallue2018,granados2018,christ2018}. Tightly packed 
orbits appear to be a natural outcome of planet formation; it is tempting to conclude 
that the Pluto-Charon satellites formed in a similar process.

In current theory, systems of planets orbiting close to resonances result from radial
migration \citep{lee2002,mustill2011,pierens2013a,pierens2013b,lin2014,macdonald2016,
mills2016,luger2017,tamayo2017,ormel2017}.
As protoplanets grow within a circumstellar gaseous disk, torques from the disk cause the
planet to migrate radially inward or outward through the disk at a rate that depends on
the mass of the planet and the physical properties of the disk. When pairs of migrating
protoplanets enter an orbital resonance, the energy required to leave the resonance is
often larger than the energy available from migration. Pairs of planets may then remain
`stuck' in the resonance or migrate together at rates that maintain the resonance. This
process can repeat for systems of planets, yielding `migration chains' where sequential
pairs of planets are at or in orbital resonance.

Applying this idea to the Pluto-Charon system requires some simple modifications. Instead
of a gaseous disk around a fairly stable, single or binary central star, newly-formed 
satellites migrate through a disk of small particles surrounding an expanding binary. 
For the observed satellite masses, typical migration rates in circumbinary disks are 
large enough to generate resonant chains of satellites \citep{kb2014b}. If the lifetime 
of the circumbinary disk is longer than the expansion time for the central binary, 
collisional damping between the small particles and the satellites is strong enough to 
maintain the resonances \citep{bk2015b}.  Once the expansion of the binary is complete, 
the satellites can disperse the disk of small particles, leaving the satellites close 
to orbital resonance as observed.

Using numerical simulations to test the plausibility of this picture requires accurate 
masses for the four small satellites. The total mass of the satellite system derived 
here, $\lesssim 10^{20}$~g, provides a new target for theoretical models that attempt
to predict the growth of satellites within circumbinary debris from the Pluto-Charon 
collision \citep{kb2014b, walsh2015} or the long-term stability of satellites as the 
binary circularizes and expands following the collision \citep{bk2015b,cheng2014b,
smullen2017,woo2018}. The smaller masses for the satellites may also improve our
understanding of their rotational evolution \citep{quill2017}. 

\vskip 6ex
\section{SUMMARY}
\label{sec: summary}

To generate robust constraints on the masses of the four small satellites
in the \pc\ system, we perform a large set of \nbody\ calculations. The
results yield clear upper limits for the masses of Nix and Hydra, $\sim$ 
10\% larger than the nominal masses derived from HST observations. 
Best estimates are $m_N \lesssim 4.7 \times 10^{19}$~g and 
$m_H \lesssim 5.5 \times 10^{19}$~g. The estimate for Nix agrees with
previous limits derived from \nbody\ simulations \citep{youdin2012}.  Results 
for Hydra are $\sim$ 50\% lower than limits derived in \citet{youdin2012}.

The calculations also demonstrate that the mass of Kerberos is much smaller 
than inferred from orbital fits to HST imaging data; $m_K \approx 10^{18}$~g 
is consistent with the \nbody\ calculations. Limits on the mass of Styx are 
much weaker; however, a plausible estimate consistent with the numerical 
calculations is $m_S \approx 6 \times 10^{17}$.

Together with large albedos derived from \nh, upper limits on the masses 
of Nix and Hydra suggest they formed in an icier mixture than Charon or 
Pluto. Using size measurements from \nh, the mass densities are
$\rho_N \lesssim$ 1.57~\gcmc\ for Nix and 
$\rho_H \lesssim$ 1.44~\gcmc\ for Hydra. A numerical analysis that 
includes uncertainties in the measured sizes implies a 70\% (80\%)
probability that the mass densities of Nix and Hydra are smaller 
than the mass density of Charon (Pluto).

These results provide new insight into the formation and evolution of 
circumbinary planet and satellite systems.  For \pc, stability of the four 
small satellites on 5--10~Gyr time scales requires satellite separations, 
$K = \delta a_{ij} / R_{H,ij} \ge$ 12, where $\delta a_{ij}$ is the absolute
value of the difference in semimajor axes between adjacent satellites and 
$R_{H,ij}$ is their mutual Hill radius. The ensemble of satellite masses 
implies formation in a moderate mass ring of material \citep[e.g.,][]{kb2014b}.

Improving constraints on the mass density of Nix and Hydra requires better 
models for their shapes and volumes. Together with new \nbody\ calculations,
frequency mapping, perturbation, and variational methods can upgrade mass 
estimates for Styx and Kerberos. Even without these advances, the new satellite
masses allow better tests of models for the formation and long-term stability
of circumbinary planets and satellites.

\vskip 6ex

Resources supporting this work on the `discover' cluster were provided
by the NASA High-End Computing (HEC) Program through the NASA Center
for Climate Simulation (NCCS) at Goddard Space Flight Center.  We thank 
M. Geller, K. Kratter, M. Payne, and A. Youdin for advice, comments, and 
encouragement.  Portions of this project were supported by the 
\textit{NASA } \textit{Outer Planets} and \textit{Emerging Worlds} 
programs through grants NNX11AM37G and NNX17AE24G.

Binary output files from the simulations, C programs capable of
reading the binary files, and some ASCII text files derived from the
simulations are available at a publicly accessible repository
(https://hive.utah.edu/) with digital object identifier TBD.

\bibliography{ms.bbl}

\begin{deluxetable}{lcccccccc}
\tablecolumns{9}
\tablewidth{0pc}
\tabletypesize{\small}
\tablenum{1}
\tablecaption{Nominal satellite properties for \nbody\ calculations\tablenotemark{a}}
\tablehead{
  \colhead{Satellite} &
  \colhead{$m_i$} &
  \colhead{$r_i$ (km)} &
  \colhead{$\rho_i$ (\gcmc)} &
  \colhead{$\rhill$ (km)} &
  \colhead{$a_i$ (km)} &
  \colhead{$e$ ($\times 10^{-3}$)} &
  \colhead{$\imath$ (deg)} &
  \colhead{$P_{orb}$ (d)}
}
\startdata
Styx (heavy) & ~~4.5 & ~~5.2 & 6.46 & 198 & 42656 & 5.787 & 0.809 & 20.16155 \\
Styx (light) & ~~0.6 & ~~5.2 & 1.02 & 101 & 42656 & 5.787 & 0.809 & 20.16155 \\
Nix & 45.0 & 19.3 & 1.49 & 487 & 48694 & 2.036 & 0.133 & 24.85463 \\
Kerberos (heavy) & 16.5 & ~~6.0 & 18.2 & 405 & 57783 & 3.280 & 0.389 & 32.16756 \\
Kerberos (light) & ~~1.0 & ~~6.0 & 1.11 & 160 & 57783 & 3.280 & 0.389 & 32.16756 \\
Hydra & 48.0 & 20.9 & 1.26 & 661 & 64738 & 5.862 & 0.242 & 38.20177\\
\enddata
\label{tab: sats}
\tablenotetext{a}{Based on published analyses of HST and \nh\ data for
the mass ($m_i$ in units of $10^{18}$ g), spherical radius ($r_i$),
mass density ($\rho_i$), Hill radius (\rhill), semimajor axis ($a_i$),
orbital eccentricity ($e$) and inclination ($\imath$), and orbital period
\citep[$P_{orb}$;][]{tholen2008,buie2013,brozovic2015,stern2015,weaver2016,
nimmo2017,mckinnon2017}
and an adopted gravitational constant $G = 6.67408 \times 10^{-8}$
\citep[see also][]{porter2015b,quill2017,kb2019}.  For Styx, the mass in 
the heavy (light) satellite system assumes a mass equal to 10\% the mass of 
Nix (a mass density of $\sim$ 1~\gcmc\ and the spherical radius from \nh\ data). 
In a heavy (light) satellite system, the mass of Kerberos is derived from 
HST data (an adopted mass density of $\sim$ 1~\gcmc\ and the spherical 
radius from \nh\ data).
}
\end{deluxetable}

\begin{deluxetable}{lcccccc}
\tablecolumns{7}
\tablewidth{0pc}
\tabletypesize{\scriptsize}
\tablenum{2}
\tablecaption{Starting conditions for $n$-body calculations}
\tablehead{
  \colhead{Satellite} &
  \colhead{$x$ (km)} &
  \colhead{$y$ (km)} &
  \colhead{$z$ (km)} &
  \colhead{$v_x$ (km~s$^{-1}$)} &
  \colhead{$v_y$ (km~s$^{-1}$)} &
  \colhead{$v_z$ (km~s$^{-1}$)}
}
\label{tab: start-pos}
\startdata
Pluto-0 & -157.9402301682 & -456.9060813269 & -2071.3208302900 & -0.0177029949 & -0.0158012994 & 0.0048365588 \\
Pluto-1 & -157.8310490659 & -456.8246643552 & -2071.4634491000 & -0.0177033115 & -0.0158016253 & 0.0048363507 \\
Pluto-2 & -157.8121679944 & -456.7988459683 & -2071.4067337364 & -0.0177032091 & -0.0158015359 & 0.0048362971 \\
\\
Charon & 1297.1743847853 &  3752.6022617472 & 17011.9058384535 & 0.1453959509 & 0.1297771902 & -0.0397230040 \\
Styx & -30572.8427772584 & -26535.8134344897 & 12311.2908958766 & 0.0232883189 & 0.0427977975 & 0.1464990284 \\
Nix & 9024.3487802378 & 15210.7370165008 & 45591.7573572213 & 0.1004334400  &  0.0865524814 & -0.0479498746 \\
Kerberos & 23564.2070250521 & 28380.0399507624 & 44578.0258218278 & 0.0792537026 & 0.0630220100 & -0.0817084451 \\
Hydra & -43331.3261132443 & -43628.4575945387 & -20506.5419357332 & -0.0374001038 & -0.0184905611 & 0.1157937283 \\
\enddata
\end{deluxetable}

\begin{deluxetable}{lcccccc}
\label{tab: density}
\tablecolumns{7}
\tablewidth{0pc}
\tabletypesize{\normalsize}
\tablenum{3}
\tablecaption{Monte Carlo Statistics for Mass Density of Nix and Hydra}
\tablehead{
  \colhead{Model$^a$} &
  \colhead{$\rho_{N,med}$} &
  \colhead{$p(\rho_N < \rho_C)$} &
  \colhead{$p(\rho_N < \rho_P)$} &
  \colhead{$\rho_{H,med}$} &
  \colhead{$p(\rho_H < \rho_C)$} &
  \colhead{$p(\rho_H < \rho_P)$}
}
\label{tab: rho}
\startdata
1a & 1.63 & 0.59 & 0.75 & 1.49 & 0.62 & 0.70 \\
2a & 1.55 & 0.68 & 0.81 & 1.29 & 0.74 & 0.79 \\
3a & 1.33 & 0.85 & 0.92 & 1.17 & 0.79 & 0.84 \\
1b & 1.61 & 0.70 & 0.92 & 1.46 & 0.78 & 0.88 \\
2b & 1.54 & 0.84 & 0.96 & 1.26 & 0.92 & 0.96 \\
3b & 1.32 & 0.95 & 0.99 & 1.14 & 0.94 & 0.97 \\
\enddata
\tablenotetext{a}{
The Monte Carlo calculations consider three models for the masses of Nix
and Hydra. Model 1: mass equal to the upper limit derived from the $n$-body
simulations, 5\% larger than the nominal mass for Nix and 15\% larger
than the nominal mass for Hydra; Model 2: gaussian distribution of mass,
with the median mass equal to the nominal mass and a dispersion equal to
3\% (4\%) of the nominal mass for Nix (Hydra); Model 3: linear distribution
of mass from a lower limit equal to 2/3 of the nominal mass to the upper
limit derived from the $n$-body simulations. For the volume, the
calculations adopt nominal semimaxes ($a$, $b$, and $c$ in km) of
$ 25 \times 17.5 \times 16$ for Nix and $32.5 \times 22.5 \times 12.5$
for Hydra and errors of (i) Model a: 2~km in each axis for Nix and
4~km in each axis for Hydra or (ii) Model b: 1~km in each axis for Nix
and 2~km in each axis for Hydra. Assuming the real sizes are normally
distributed about the nominal sizes, the adopted volume is
$V = 4 \pi a b c /3$. For any realization of mass and volume in
the Monte Carlo calculation, the mass density is $\rho = m / V$.
The probabilities listed in columns 3, 4, 6, and 7 denote the fraction
of 10001 realizations with mass density smaller than the measured mass
density of Charon or Pluto. Median densities in units of g~cm$^{-3}$
are in columns 2 and 5.
}
\end{deluxetable}

\end{document}